# Virtual Net: a Decentralized Architecture for Interaction in Mobile Virtual Worlds

Bingqing Shen[1] and Jingzhi Guo[1]

[1] Faculty of Science and Technology, University of Macau, Macau, China.

Correspondence should be addressed to Jingzhi Guo; jzguo@umac.mo

## Abstract

With the development of mobile technology, mobile virtual worlds have attracted massive users. To improve scalability, a peer-to-peer virtual world provides the solution to accommodate more users without increasing hardware investment. In mobile settings, however, existing P2P solutions are not applicable due to the unreliability of mobile devices and the instability of mobile networks. To address the issue, a novel infrastructure model, called Virtual Net, is proposed to provide fault-tolerance in managing user content and object state. In this paper, the key problem, namely object state update, is resolved to maintain state consistency and high interaction responsiveness. This work is important in implementing a scalable mobile virtual world.

## 1. Introduction

Virtual worlds, including multiplayer online games and virtual social worlds, allow users to inhabit in virtual environments, create their own content, and interact with each other. Mobile virtual worlds allow users to access the simulated environments through mobile devices, achieving the possibility to play anywhere. Mobile virtual worlds have gained large attraction from the development of mobile devices. They have become an important market and revenue source for the game industry, and attracted a large number of users. For example, Fortnite has earned \$1,996,917 gross daily revenue[1] and reported 3.4 million concurrent players[2] in 2018. The success and expansion of mobile virtual worlds raise new challenges in infrastructure development, one of them is the scalability problem. In virtual worlds, interaction is implemented by sending events to servers for processing and receiving updates from the servers for rendering and state synchronization. With the increase of concurrent online users, more computing load is imposed on game infrastructures. Servers have to process and respond to more client requests within a short period for high responsiveness. Also, network bandwidth consumption is increased to pack multiple game states in an update. For scaling, more computing resources have to be invested. Otherwise, user experience will be affected.

Peer-to-peer (P2P) virtual worlds, firstly introduced in [1], explore the possibility of running a virtual world without a central server. In P2P virtual worlds, user devices run both the client program and server program for event handling and state update. Thus, computing resources naturally scale along with the change of user population. Mobile applications, however, have different characteristics with respect to their desktop counterparts. One outstanding issue is client failure. Compared to desktop PCs, mobile devices are more prone to failure, due to, for example, battery depletion or application conflict. Moreover, the access to mobile networks, such are MANETs and VANETs, are also unstable. Client unreliability may cause content loss or state inconsistency, if user content and object state are not properly saved or backed up before failure. Yet, existing P2P virtual worlds do not concern the peer device unreliability problem [3]. Thus, they cannot be directly applied in mobile settings.

---

[1] Thinking Gaming (2018, July), Top grossing iPhone - Games. https://thinkgaming.com/app-sales-data/
[2] Metro (2018, July), Fortnite overtakes PUBG as the biggest video game in the world. https://metro.co.uk/2018/02/09/fortnite-overtakes-pubg-biggest-video-game-world-7300442/

In this paper, a Virtual Net model is proposed to address the client unreliability problem for mobile P2P virtual worlds. The model utilizes the cloud-fog structure, but totally decentralized. To avoid content loss, the cloud layer stores user contents for content persistency. The fog layer caches object states for client recovery and maintains state consistency. The separation of content storage and state caching can improve responsiveness, since operations direct on P2P storage have more communication overhead [21]. Based on the P2P content storage, a content addressing scheme is devised, which can facilitate content integrity check.

To avoid reinventing the wheel, this paper mainly focuses on the state update problem to maintain object state consistency. At the fog layer, object states are replicated on several nodes for fault-tolerance. Thus, all replicas must maintain the same state in event handling so that interaction can be performed within a consistent shared environment. Yet, the requirement of high responsiveness in virtual world interaction makes the problem difficult. To attack the difficulty, an opportunistic approach, called fast event delivery, is proposed. Based on the approach, a virtual world interaction model is then designed. In short, the main contributions of the paper are listed as follows.

1. A new P2P cloud-fog structure, called Virtual Net model, is proposed to resolve the client unreliability problem, which can provide fault-tolerance in playing a mobile virtual world.
2. A fast event delivery approach is proposed to both maintain replica state consistency and high responsiveness in the process of handling user events.
3. A new virtual world interaction model is designed to achieve game state consistency and high responsiveness when interacting with different neighbors.

The remainder of the paper is organized as follows. The related works are introduced in Section 2. The overall Virtual Net model is described in Section 3. Section 4 studies the state update problem in detail. Based on the solution of the problem, the virtual world interaction model is provided in Section 5 with neighbor change management. The correctness of the solution is proved in Section 6. Section 7 and Section 8 evaluate the performance through theoretical analysis and experiments. Section 9 concludes the paper.

## 2. Related Work

Mobile P2P virtual worlds combine the characteristics of mobile virtual world and P2P virtual world problems. Due to the lack of study in this field, the related work in P2P virtual worlds and cloud-fog mobile applications are surveyed to shape the distinct characteristics of the combined problem.

## 2.1. P2P Virtual Worlds

P2P MMORPGs and P2P virtual environments have been amply surveyed in [3] and [4]. Previous works mainly focus on inter-player consistency management, including peer connectivity, interest management, event dissemination, and cheat prevention. Peer connectivity [5] studies the connection of all user devices within an overlay network such that any peer can be reached from another peer. Interest management [2] restricts the range of message receipt to reduce communication overhead in state update. Event dissemination [6] reduces the number of communication channels on event senders to avoid overwhelming them in hotspot areas. Cheat prevention [7] is needed to achieve fairness without the arbitration from a central server. In these work, a desktop environment is assumed such that a client is always reliable in storage and connection. In contrast, the Virtual Net targets at the mobile environments in which both devices and connections are unreliable, which is the new problem and orthogonal to the above studies. Thus, a complete implementation of Virtual Net can employ existing P2P solutions in inter-player consistency management, such as peer connectivity and interest management, to avoid reinventing the wheel.

Early work on P2P state persistency is related to the content storage in this work. State persistency studies the reliable storage and efficient retrieval of user state [8]. Each time a state is updated, it has to be persisted in the overlay network, and the state has to be queried from the overlay network when the client is recovered from a failure. Same as the above argument, the work in [8] only assume a reliable client, which is not applicable in a mobile setting. The Virtual Net model not only solves the unreliable client problem, but also reduces storage and retrieval overhead through content caching. Moreover, content integrity check is included in Virtual Net, which is not mentioned in previous works.

## 2.2. Fog Computing

Firstly introduced in [9], cloud gaming moves the game engine functions to the cloud to simplify development, distribution, access, and update [10]. However, the measurement study [11] shows that the current cloud gaming infrastructure is unable to meet the latency requirement for end-users distant from data centers. To improve latency, fog computing [12] has been introduced to move the time-critical functions to the locations near clients. Fog computing has been widely discussed in both Internet of Things (IoT) [13] and mobile computing [14] to offload server burden [15], enable location awareness, and provide real-time interaction. Among its many applications, mobile gaming [16] and mobile reality [17] are two important examples. Similar to the cloud-fog structure, the Virtual Net solution also employs the cloud layer for content storage and the fog layer for latency improvement. But differently, Virtual Net explores a totally decentralized solution, with no central control at the cloud layer.

# 3. Virtual Net Model

The proposed Virtual Net structure is based on the commonly used three-layer structure shown in Figure 1. Similar to some existing cloud-fog structures, it is divided into three layers: the cloud layer (L1), the fog layer (L2), and the client layer (L3). The cloud layer provides persistency service, which stores the files of user content and the state of virtual objects (avatars, accessories, achievements, etc.). The fog layer caches object states in play and provides state recovery for clients in case of short-term failure. It also periodically checks object states and saves them to the cloud layer for state persistency, which is asynchronous to event handling. When a user leaves a game, the cached state of user object will be saved to the cloud layer. The client layer provides user interfaces for receiving user operations and displaying updated states for user interaction. Virtual worlds are latency-sensitive applications. Yet, on one hand, clients require fast state update in user interaction [18]. On the other hand, the complexity of peer-to-peer routing slows down the process of content storage and retrieval [19]. Thus, the fog layer is padded between L1 and L3 to improve responsiveness in fault-tolerance.

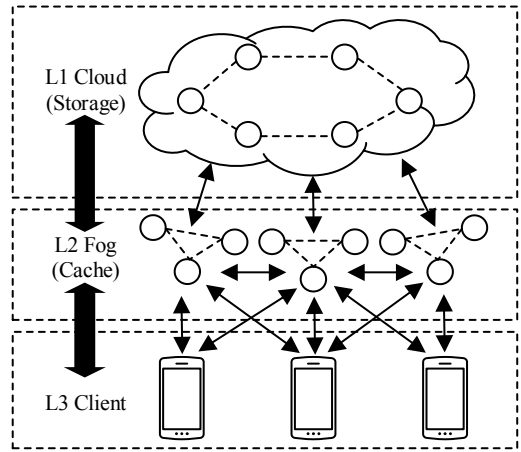

Figure 1: Virtual Net Overall Model

The three-layer architecture is resilient. First, L1 and L2 can be individually scaled without affecting each other, since they are built for different purposes. The cloud layer focuses on the long-term storage of user content, which is only accessed at user login, logout, and periodic state checkpoint by the fog nodes. On the other hand, the fog layer maintains the latest state of user content and provides state recovery from intermittent client failure. Except for state initialization and checkpoint, L1 and L2 do not need to interact with each other. Besides, the model provides some extent of isolation of failure. The failure of one layer can be recovered by another layer, since each layer has a separate copy of content.

Different from the existing cloud-fog computing paradigm, computing resources in the cloud and fog layer are P2P nodes, like BitTorrent or eDonkey. Specifically, users contribute part of the computing resources from their devices which can be smartphones, laptops, desktop PCs, or even servers. A device is divided into one or several virtual nodes [20] for fine-grained load balancing. All virtual nodes are managed by a node pool. For different computing purposes, there are two types of virtual nodes: storage nodes and cache nodes. The storage nodes construct the cloud layer and the cache nodes construct the fog layer. Thus, Virtual Net is a decentralized computing paradigm. A client could be on the same device of a virtual node, like BitTorrent, or on a separate lightweight device.

### 3.1. P2P Cloud Layer

Object files are stored on the cloud layer through P2P file storage. Based on the file storage system, a content addressing scheme is devised, which can not only provide flexibility in content identification and addressing but also provide integrity in object management.

### 3.1.1. File Storage

The TotalRecall [21] storage architecture is applied to manage the storage nodes for file storage. The details of the design and performance can be found in [21]. Here, only the overall mechanism is introduced. In TotalRecall, each node is assigned a unique hash code as the node ID. Also, each file has a file ID which is the hash checksum of the file. When a new file is created, the file is associated to a storage node, called the master node whose ID is closest to the file ID. Other nodes hosting the data of the file are called host nodes. Master nodes manage the location of host nodes and the version control for the associated files. Each storage node can be the master node for some files and the host node for other files. Thus, the entire storage node network forms a distributed hash table (DHT) for file lookup. To request a file, its master node is found first with the file ID. Then, based on the reply from the master node, the host nodes are located and the file can be retrieved (or reconstructed).

### 3.1.2. Content Addressing

To retrieve the objects from the cloud layer, object content needs to be identified and addressed. A hierarchical content addressing scheme is devised, which can facilitate content integrity check. The devised content addressing scheme has four hierarchies: inventory, objects, components, and files, as illustrated in Figure 2.

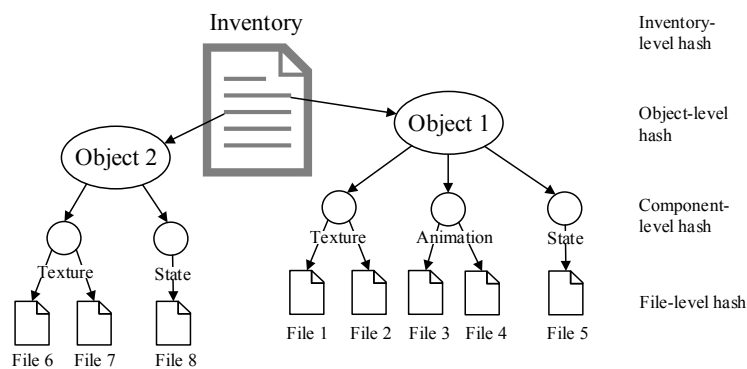

Figure 2 Illustration of the content addressing scheme hierarchy.

(*Inventory-level*) Each user has an inventory file, identified by the inventory ID which is the hash code of the user ID. An inventory contains all the object descriptions, consistently managing content identification and modification. Thus, to retrieve the object contents, the inventory file needs to be retrieved first. (*Object-level*) An object is identified by the object hash code and composed of one or multiple components. (*Component-level*) Each component is identified by the component hash code. Object components are the categories of object resource files, which are classified into animation, sound, texture, script, etc. (*File-level*) The actual files of objects are addressed by file IDs in object descriptions. Through files ID, the actual file can be retrieved either from the local cache or from the DHT of the cloud storage.

Based on the structure of the content addressing scheme, a Merkle tree [22] (Figure 2) can be hierarchically constructed with the file hash code, component hash code, object hash code, and inventory hash code. With the Merkle tree, the integrity of user content can be recursively checked and the number of hash comparison can be largely reduced [23]. Typically, a client caches more than 500,000 files of user-created contents [24]. Thus, an exhaustive search of updated files will be inefficient. We conduct an experiment with 200 objects and more than 5,000 files. Compared with the file-level and object-level content integrity check [24], Figure 3 shows that the proposed four-level content integrity verification has fewer hash comparisons, especially with respect to a small number of file changes.

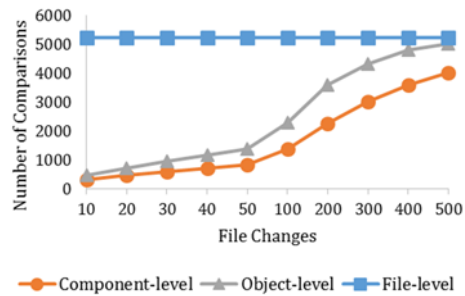

Figure 3 Number of hash comparisons with random file changes in content integrity check.

## 3.2. P2P Fog Layer

The fog layer is added between the cloud layer and the client layer to mask the latency of content storage, and meanwhile provide fault-tolerance. From the user perspective, each user is allocated some cache nodes, when he/she is playing in a virtual world. These cache nodes provide the user some computing resources, forming a logical computing unit. We call it mesh computer, as illustrated in Figure 4. When a user logs to the system, her client firstly initializes the mesh computer by requesting for some cache nodes from the node pool. The cache nodes then retrieve the content from the cloud layer. The client also retrieves the saved content from the cloud layer for content rendering and state synchronization. When the mesh computer receives a quit instruction from the client or the client is experiencing a long-term failure, the mesh computer will release the cache nodes to the node pool. Optimal resource allocation and cost minimization have been studied in [20], which is out of the scope of this paper.

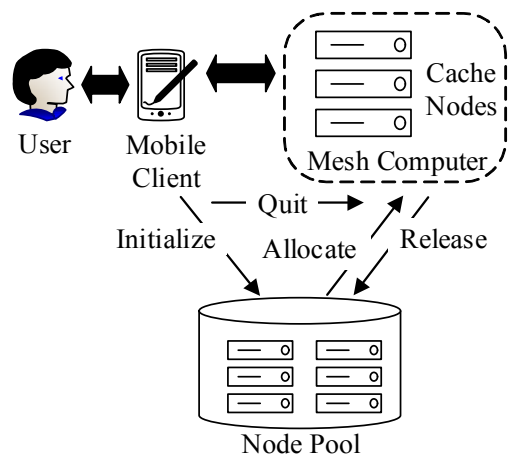

Figure 4: Mesh computer.

Due to the unreliability characteristic of P2P nodes, they are subject to (either temporarily or permanently) failure. Thus, for reliability purpose, a mesh computer maintains multiple cache nodes which are the replicas of the same user content, called a replica group. Content will be transferred from failed nodes to live nodes. Replica group management has been studied in our previous work [25]. This paper focuses on replica state management in the following sections.

# 4. Object State Update

At the fog layer, it is important that all replicas of the same group maintain the same state of user objects so that any failure of a replica will not invalidate a user's current state. The problem becomes challenging, since replicas could receive different sets of concurrent events from different senders and events could be received in different orders. The state machine replication (SMR) [26] approach is adopted to manage object state. SMR is a fault-tolerance model replicating a deterministic finite state machine on a set of distributed nodes, each of which has the same input, output, and state transfer. In an asynchronous cycle, firstly, clients send requests to all the nodes. On receiving the requests, a consensus protocol is triggered to determine the sequence of requests. Then, all nodes process the requests in the decided sequence so that they can reach the same new state. To reduce communication overhead, one replica is elected as the leader, coordinating membership reconfiguration and request ordering.

In virtual worlds, however, the consensus process adds a large delay in user interaction, because a requested event must be agreed by all replicas after at least two communication rounds (i.e., four communication steps) [27] to reach an agreement, before it can be handled and replied to clients. The interaction delay issue in event handling is addressed based on the following observations. Due to users' limited perception range and motion speed, the number of event senders within a small period is fixed. Thus, the number of concurrent event senders within the period can be known a priori. Based on this observation, we propose a fast event delivery approach.

## 4.1. Fast Event Delivery

Fast event delivery allows a replica to directly deliver a received event through a cycle-event mapping, if it can ensure that the same event will eventually be delivered by all replicas. Specifically, the timeline is divided into infinite cycles of length $\Delta t$. From cycle $c_0$, an event sender $s$ periodically broadcasts an event to the replicas in each cycle. Each event is identified by the sender ID and the sequence number. The sequence number of the first event at cycle $c_0$ is 0. If there is no operation, $s$ just broadcasts a no-op event. At the receiving end, $c_0$ is also known by all replicas. Each replica delivers an event with sequence number $c - c_0$ from $s$ for cycle $c$, which is called the event of cycle $c$. Events for cycle $c$ from different senders will be ordered by sender ID. If a replica does not receive the event for cycle $c$ from $s$, it will start an instance of consensus for the cycle. In the consensus, if a replica has received the event for cycle $c$, that event will be decided by the leader and delivered by all replicas. Otherwise, they will decide and deliver an empty event for cycle $c$. Events will be delivered to a queue $Q_d$ first and then sent to the application from the queue for handling in sequence.

The relation of cycle ($c_i$), sender ID, event sequence number, delivery queue ($Q_d$), and delivery sequence ($\lambda$) are illustrated in Figure 5. Specifically, in $Q_d$, the subscript of event $e$ denotes the event sequence number which is equal to the event sending cycle. Thus, for the same cycle, the events in $Q_d$ are sorted by the sender ID and mapped to the local index numbers in $Q_d$ (i.e., the second member in the tuples of $Q_d$). $\lambda$ represents the global index of events delivered to the application, which will be introduced in Section 4.1.4.

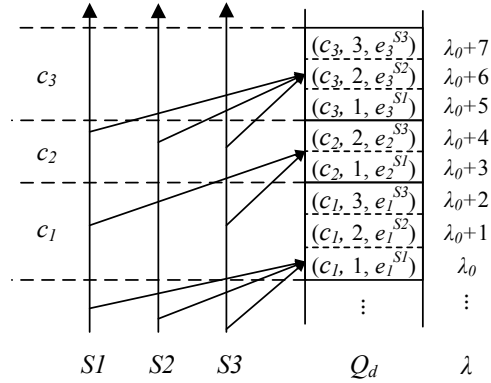

Figure 5 Relation of cycle ($C_i$), sender ID ($Si$), event sequence number, delivery queue ($Q_d$), and delivery sequence ($\lambda$).

Figure 6 illustrates the fast event delivery process with one sender $s$ and three replicas $r_1$, $r_2$, and $r_3$ in the replica group $g$. $s$ broadcasted four events, $e_1$, $e_2$, $e_3$, and $e_4$, at cycle $c_1$, $c_2$, $c_3$, and $c_4$ to $g$. All replicas received $e_1$ at cycle $c_2$ and $e_4$ at $c_5$. Only replica $r_1$ received $e_2$ at $c_3$. No replica received $e_3$ at $c_4$, but $r_1$ and $r_2$ received $e_3$ at $c_5$. $r_1$, $r_2$, and $r_3$ deliver $e_1$ for $c_1$. They then collectively decide $e_2$ for $c_3$ and an empty event for $c_4$ through consensus. The first problem is how to decide $c_5$ and $e_3$. According to the cycle-event mapping principle (i.e., one-cycle-one-event), the replicas should only deliver $e_4$ for $c_5$ and discard $e_3$, leading to event loss. Before discussing the late events handling problem in detail, the settings and assumptions of the system will be introduced first.

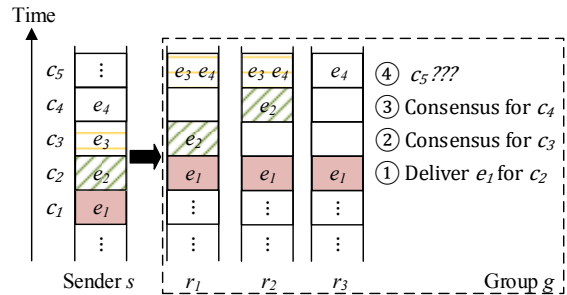

Figure 6: Illustration of fast event delivery.

### 4.1.1. Settings and Assumptions

For fault-tolerance, a replica group contains at least $n$ replicas. The minimal group size $n$ is determined by the content availability requirement [25] and the replica failure rate. To reduce replication overhead, each group also has an extra number $e$ of nodes for lazy repair [21]. Once $e + 1$ replicas fail, new replicas will be added to recover the group size to $n + e$.

In each replica group, there is one non-replica node monitoring the state of all replicas, called Rendezvous [28]. A Rendezvous uses timeout to determine the state of replicas, and then broadcasts their states to all replicas. Monitoring replica state is implemented by exchanging heartbeat messages between a replica and a Rendezvous. If the Rendezvous does not receive one heartbeat message within a cycle, the replica is treated as failed and removed from the group. New replicas are also added by the Rendezvous, once the group size is smaller than $n$. Rendezvouses are reliable nodes, or called super-peers [16], since the existence of a group is determined by the Rendezvous. Once a Rendezvous fails, a new Rendezvous must be assigned to the replica group, which then rebuilds the replica group and recovers the object states from the cloud storage. By exchanging heartbeat messages, each replica learns the current membership of the group $g$, denoted by $G$, which contains all live replicas of group $g$. When the Rendezvous tells a replica that a member has failed or a new member is added, the replica will remove the member from $G$ or add the member into $G$.

The system is assumed to be live. The SMR model contains three types of group-wide activities: leader election, group reconfiguration, and consensus. The liveness assumption ensures that, when an activity is needed, it will eventually succeed after a finite number of failures.

Each replica group maintains a set of event senders. It is assumed that each event sender has an ID which is globally unique and sender IDs are comparable. Moreover, a replica group will append the join timestamp to sender IDs to distinguish a sender in two different joins of a sender set.

Let $s$ be the ID of an event sender, $c$ be a cycle number of a replica group $g$, $r_i$ be the $i^{th}$ replica in group $g$. Some important relations of event, event sender, event sequence number, and cycle number are defined in Table 1. Other notations used throughout the subsequent sections are listed in Table 1. Besides, event names are capitalized.

Table 1 Notations

| Notations | Descriptions |
|---|---|
| Relations of event, event sequence number, sender, and cycle number | |
| $Seq(s, c)$ | The sequence number of the event sent for cycle $c$ from sender $s$ |
| $Seq(e)$ | The sequence number of event $e$ |
| $e(s, j)$ | The event of sequence number $j$ sent from $s$ |
| $Deliver(e, c)$ | Return $true$ if event $e$ is deliverable at cycle $c$. Otherwise, return $false$ |
| $DeliverCycle(e)$ | Return the cycle in which event $e$ is delivered. If $e$ has not been delivered, return $\perp$. |
| $Receive(e, c)$ | Return $true$ if $e$ has been received at cycle $c$. Otherwise, return $false$. |
| $ReceiveCycle(e)$ | Return the cycle in which event $e$ is received. If $e$ has not been received, return $\perp$. |
| $Event\ of\ cycle$ | The event sent by a sender for a cycle is called the event of the cycle. Given the first event sent from sender $s$ at cycle $c_0$. The event of cycle $c$ satisfies: $Seq(e) = Seq(s, c) = (c - c_0)$. |
| $Cycle\ open\ /\ close$ | If a replica $r_i$ has delivered the events for cycle $c$ and moved to the next cycle $c + 1$, then $c$ is closed and $c + 1$ is still open to $r_i$. |
| Other notations | |
| $r_i$ | Replica $i$ |
| $r_L$ | Group leader |
| $G$ | The set of group members |
| $S$ | The set of event senders |
| $U$ | Set of update recipients |
| $g$ | Replica group $g$ |
| $s$ | Sender $s \in S$ |
| $Index(s)$ | The index of $s$ in $S$ sorted by sender ID |
| $Sender(e)$ | The sender of event $e$ |

| $c$ | Cycle number $c$ |
|---|---|
| $\Delta t$ | Cycle length |
| $t_{start,s}$ | The start time of the first cycle for $s$ |
| $t_{send,s}(n)$ | The time to send the $n^{th}$ event from $s$ |
| $t_{recv,s}(n)$ | The time to receive the $n^{th}$ event sent from $s$ |
| $t_{now}$ | Current time |
| $E / E(c)$ | The set of decided events / decided events of $c$ |
| $Q_d$ | Delivery queue, containing the sorted events to be delivered to the application in sequence |
| $(c, \gamma, e)$ | The event $e$ delivered for cycle $c$ with local index $\gamma$ in $Q_d$ |
| $(\lambda, e)$ | The delivery sequence $\lambda$ of event $e$ in $Q_d$, one-to-one mapped onto $(c, \gamma, e)$ for the given $e$. |
| $\Omega(s, c)$ | The set of expected events which can be delivered to $Q_d$ (called deliverable events) from $s$ at cycle $c$ |
| $\Pi(s, c)$ | The set of actual received events from $s$ at cycle $c$ |

### 4.1.2. Late Event Handling

An event is late if the event of cycle $c$ is received after $c$ on all replicas. Formally, a late event $e$ satisfies: $Seq(e) = Seq(s, c) \land ReceiveCycle(e) > c$ on $r_i \in G$ and $Seq(e) = Seq(s, c) \land (ReceiveCycle(e) > c \lor ReceiveCycle(e) = \perp)$ on $r_j \in G \setminus \{r_i\}$. For example, $e_3$ in Figure 6 is a late event.

To ensure the agreement of cycle event delivery on all replicas, a late event can be simply discarded, since any event can be re-sent by a client with a new sequence number if the client does not receive the reply for the event for a period. However, if a sender's clock is temporarily out-of-sync with the replicas' clock or a large sending delay is experienced, a large number of events could be discarded and need re-send, as shown in Figure 7.

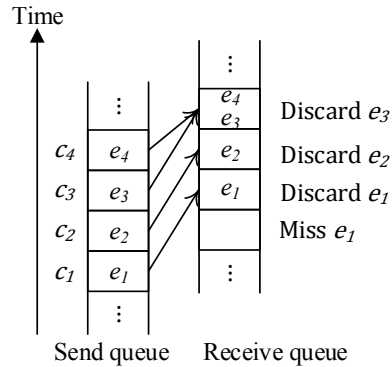

Figure 7: An extreme scenario of late event discard.

To address the late event handling problem, a dynamic cycle event delivery approach is proposed, which includes two conditions for late event delivery. The purpose of the approach is to minimize the number of event discard, and meanwhile each replica can decide the delivery of late events with only local information. Below are the conditions of late event delivery. In short, only late and out-of-order events will be discarded.

1. At cycle $c$, all events from sender $s$ with sequence number less than $c$ will be deliverable. Formally, $\forall e(s, j_1), e(s, j_2), ..., e(s, j_n)$,

$$\left.\begin{array}{c} Receive(e(s,j_1),c) \wedge \cdots Receive(e(s,j_n),c) \\ \wedge \big(DeliverCycle(e(s,j_1)) = \cdots = DeliverCycle(e(s,j_n)) = \perp\big) \\ \wedge \big(seq(s,j_1) < seq(s,c) \cdots seq(s,j_n) < seq(s,c)\big) \end{array}\right\} \to Deliver(e(s,j_1),c) \wedge \cdots \wedge Deliver(e(s,j_n),c).$$

2. At cycle $c$, an event will be non-deliverable, if one of its subsequent events has been delivered before $c$. Such event is a late and out-of-order event. Formally, $\forall e(s, j)$,

$$\left.\begin{array}{c} Receive(e(s,j),c) \\ \wedge (\exists e(s,k) \wedge k > j) \\ \wedge \big(DeliverCycle(e(s,j)) = \perp\big) \\ \wedge \big(DeliverCycle(e(s,k)) < c\big) \end{array}\right\} \to \neg Deliver(e(s,j),c).$$

To implement the dynamic cycle event delivery approach, the lowest deliverable sequence number from any sender needs to be determined first for all cycles. Specifically, at cycle $c$, let $MinSeq(s, c)$ be the lowest sequence number of all undelivered events from sender $s$ and $MinSeq(s, c) \leq seq(s, c)$. Also, let $MaxSeq(s, c)$ be the sequence number of the last delivered non-empty event in cycle $c$ from $s$. Then, $MinSeq(s, c) = MaxSeq(s, c-1) + 1$, where $MaxSeq(i, c-1)$ is determined by the event delivery for cycle $c-1$. Define the set of expected deliverable events from $s$ at cycle $c$ by $\Omega(s, c) = [MinSeq(s, c), Seq(s, c)]$ and the set of actual received events $\Pi(s, c)$. The actual deliverable events from $s$ at cycle $c$ can be filtered by $\Omega(s, c) \cap \Pi(s, c)$, which excludes the late and out-of-order events.

### 4.1.3. Total-order Event Delivery

Total-order event delivery is the key mechanism in object state update to ensure that all replicas in the same group can reach the same state along the same path of state transfer, if no more event is received. By applying the dynamic cycle event delivery approach, the event delivery for one cycle is described in Algorithm 1, where $E(c)$ stores the events from the consensus for cycle $c$ and $\gamma$ is the event delivery index in cycle c. $\gamma$ is calculated by $\gamma = Seq(e) + \sum_{k=1}^{Index(Sender(e))-1}(Seq(k,c)+1)$. $(c, \gamma)$ and the calculation of $\gamma$ ensures that the events in $Q_d$ are sorted first by cycle number, then by sender index, and lastly by event sequence number, which can sort all events in the same order on all replicas.

In a run for cycle $c$, each replica firstly checks whether there are any events decided for the cycle from any consensus instance. If they exist, these events will be directly moved to $Q_d$ for event handling by the application. Otherwise, Algorithm 1 checks the condition $\Pi(s, c) = \Omega(s, c)$ for each sender $s$ to ensure whether all expected deliverable events have been received. If there is an expected deliverable event not received in this cycle, then the replica will trigger a consensus instance to determine the event delivery for cycle $c$. Note that the cycle number $c$ will only be increased if the events of the cycle have been delivered.

| Algorithm 1 Event Delivery |
| --- |
| 1.  For cycle $c$, |
| 2.  If $E(c) \neq \emptyset$, then |
| 3.      $Q_d \leftarrow Q_d \cup \{(c, \gamma, e(s, j)) \mid (c, e(s, j)) \in E(c)\}$ |
| 4.      $c \leftarrow c + 1$ |
| 5.  Else, |

| | |
|---|---|
| 6. | For $\forall s \in S$, |
| 7. | If $\Pi(s, c) = \Omega(s, c)$, then |
| 8. | $Q_t \leftarrow Q_t \cup \{(c, \gamma, e(s, j)) \mid j \in \Omega(s, c)\}$ |
| 9. | Else, |
| 10. | $Q_t \leftarrow \emptyset$ |
| 11. | Consensus for $(c, \bigcup_{i=0}^{|S|-1} \Omega(s, c))$ |
| 12. | End the loop |
| 13. | If $Q_t \neq \emptyset$, then |
| 14. | $Q_d \leftarrow Q_d \cup Q_t$ |
| *15.* | *$c \leftarrow c + 1$* |

The proposed consensus algorithm is described in Algorithm 2 and illustrated in Figure 8. The consensus request is composed of the cycle number $c$ and the sequence number of all the expected deliverable events in $c$ from all senders. By receiving the consensus request, each replica proposes the actual deliverable events to the leader. If a replica does not receive an expected event, it will propose $\bot$ for the event. On receiving all proposals, the leader then decides the event for each sender and each expected sequence number. If at least one replica proposes a non-$\bot$ and non-empty event $e(s, j)$ for $j \in \Omega(s, c)$, then $e(s, j)$ will be decided for sequence number $j$ from $s$. Otherwise, an empty event will be decided for the slot. After the events of all slots have been decided, the leader will broadcast the decision to all replicas, and they will move the decided events to $E(c)$ for event delivery after receiving the decision. It is assumed that reliable point-to-point and multicast communication channels [27] are applied in the consensus protocol.

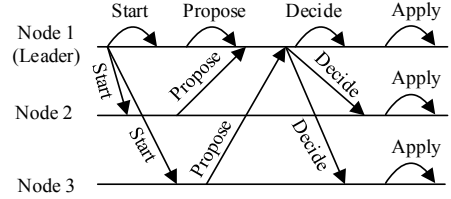

Figure 8: Illustration of the consensus protocol.

| Algorithm 2 Consensus |
|---|
| 1.   Given $(c, \bigcup_{i=0}^{|S|-1} \Omega(s, c))$, |
| 2.   $r_i$ proposes: $\{(s, j, e(s, j)) \mid s \in \boldsymbol{s} \wedge j \in \Omega(s, c) \cap \Pi(s, c)\}$ |
| 3. |
| 4.   $r_L$ decides: |
| 5.       For each $s \in \boldsymbol{s}$ and $j \in \Omega(s, c)$, |
| 6.           // Proposals: the set of proposed events for $c$ from all replicas; |
| 7.           If $\nexists e(s, j) \neq \bot \wedge e(s, j) \in$ Proposals, then |
| 8.               $e(s, j) \leftarrow Empty$ |
| 9.           $D(c) \leftarrow D(c) \cup \{(c, e(s, j))\}$ |
| 10. |
| 11.   $r_n$ applies the decision: $E(c) \leftarrow D(c)$ |

The complete algorithms of total-order event delivery, including event collection, event delivery, and consensus can be found in Appendix A.

### 4.1.4. Garbage Collection

To avoid buffer overflow, events which have been delivered to the application need to be removed from $Q_d$ to avoid $Q_d$ from unlimited growth or even overflow. Due to asynchronous event handling, however, a replica $r_i$ cannot safely remove a delivered event only with local information, because another replica $r_j$ later may request for the removed event for a given cycle if $r_j$ does not receive it. Thus, determining the removable events is a challenging problem.

Learned from Algorithm 1, since all events in $Q_d$ can be uniquely identified by $(c, \gamma)$, there exists a relation that maps each $(c, \gamma)$ to a unique integer number $\lambda \in [0, \infty)$, called delivery sequence. Thus, events in $Q_d$ can also be identified by $(\lambda, e)$ and each $(c, \gamma, e)$ and $(\lambda, e)$ has a one-to-one mapping for the same $e$. Let $Q_d$ be a sequence $e_i e_{i+1}... e_j...e_k$ mapped to $\lambda_i \lambda_{i+1}... \lambda_j... \lambda_k$. It can be observed that the events which can be safely pruned satisfy the following characteristics.

1) If $e_j$ can be pruned from $Q_d$, then all events before $e_j$ can all be pruned from $Q_d$.
2) Let $L = e_i e_{i+1}...e_j$ be the sub-sequence of $Q_d$. If $L$ can be pruned from the $Q_d$ on one replica, then $L$ can be pruned from the $Q_d$ of all the replicas in group $g$.
3) If $e_j$ can be pruned from $Q_d$, then $\lambda_j \leq \lambda_c$, where $\lambda_c$ is the sequence of the last event handled by the application, returned by the function $LastApplied(Q_d)$.
4) Trailing rounds with empty events cannot be removed from $Q_d$, since the events of these rounds may be used in consensus for deciding the value of late events.

1) and 2) implies that there is a common latest applied event $e_{cle}$ such that all the events delivered before $e_{cle}$ (including $e_{cle}$) have been applied on all the replicas, whereas the events after $e_{cle}$ are undecidable. The delivery sequence of $e_{cle}$ is denoted by $\lambda_{cle}$. 3) implies that the $\lambda_{cle}$ cannot exceed $\lambda_c$. 4) restricts the range of garbage collection. Thus, all the events before $\lambda_{cle}$, which are not empty trailing events, can be safely removed from $Q_d$.

Based on the above observation, a gossip protocol is devised to learn $\lambda_{cle}$ by exchanging $\lambda_c$ of all replicas for safely estimating the earliest removable event, which is described in Algorithm 3. In the protocol, each replica periodically sends its $\lambda_c$ to other replicas. A replica also caches the received $\lambda_c$. Based on the latest received $\lambda_c$ from all replicas, $\lambda_{cle}$ can be determined by the minimal $\lambda_c$. Then, $\lambda_{cle}$ is adjusted to exclude the trailing empty events (Line 12-16). Lastly, all the events before $\lambda_{cle}$ are removed from $Q_d$. In Algorithm 3, $\Lambda$ caches the received $\lambda_c$'s from all replicas in $G$. For a $\lambda_c$ from $r_i$, if $\lambda_c$ is greater than the cached value of $r_i$ in $\Lambda$, the new $\lambda_c$ can safely replace the existing one since $\lambda_c$ from the same replica are monotonically non-decreasing.

| Algorithm 3 Garbage Collection Protocol |
|---|
| 1.    On replica $r_i$: |
| 2.    Upon Timer TIMEOUT |
| 3.       $\lambda_c \leftarrow LastApplied(Q_d)$ |
| 4.       Broadcast $\lambda_c$ to all $r \in G$ |
| 5.       Reset Timer |
| 6. |
| 7.    On replica $r_j$: |
| 8.    Upon $\lambda_c$ from $r_i \wedge \lambda_c > \Lambda(r_i)$ |
| 9.       $\Lambda \leftarrow \Lambda \cup \{(r_i, \lambda_c)\}$ |
| 10.      If $|\Lambda| \geq |G|$, then |
| 11.         $\lambda_{cle} \leftarrow Min\{ \lambda_c \mid (r_i, \lambda_c) \in \Lambda \}$ |
| 12.         If $(\lambda_{cle}, e_{cle}) = Q_d.last$   // last event in $Q_d$ |

| 13. | Map $\lambda_{cle}$ to $(c, \gamma)$ |
| 14. | While $\exists (c, \gamma, e) \in Q_d \wedge e = Empty$ |
| 15. | $\lambda_{cle} \leftarrow \lambda_{cle} - |\{(c, \gamma, e) | (c, \gamma, e) \in Q_d\}|$ |
| 16. | $c \leftarrow c - 1$ |
| 17. | $Q_d \leftarrow Q_d \setminus \{(\lambda, e) | \lambda \leq \lambda_{cle}\}$ |

*4.1.5. Time Synchronization*

In the fast event delivery approach, another key component is the synchronization of the start and end time of a cycle on event senders and recipients (all the replicas in group g) to minimize the chance of handling late events through consensus. Specifically, let $\Delta t_n$ be the amount of network latency. Assume that the upper bound and lower bound of $\Delta t_n$, denoted by $\Delta T_n{}^L$ and $\Delta T_n{}^H$, can be estimated such that most $\Delta t_n$ falls within the range $[\Delta T_n{}^L, \Delta T_n{}^H]$. Then, cycle length $\Delta t$ can be determined by $\Delta t = (\Delta T_n{}^H - \Delta T_n{}^L)$.

Let $t_{start,s}$ be the start time of the first cycle for event sender $s$ designated by the replicas. Also, let $t_{send,s}(n)$ be the send time of the $n^{th}$ event from $s$ to the replicas. $s$ firstly calculates $t_{send,s}(1)$ by $t_{send,s}(1) \leq t_{start,s} - \Delta T_n{}^L$. Then, in the $n^{th}$ cycle, it calculates the sending time of the $n^{th}$ event by $t_{send,s}(n) = t_{send,s}(1) + (n - 1) \cdot \Delta t$. At the receiving end, all replicas are timed to receive the $n^{th}$ event from $s$ at $t_{recv,s}(n) = t_{send,s}(n) + \Delta t = t_{start,s} + n \cdot \Delta t$. To timely collect received events, event collection and event delivery can be run by different threads with different buffers (See Appendix A for the details).

A time server, such as a NTP server [29], can be deployed to the system for synchronizing the clock of event senders and recipients, which can improve the performance of the system.

## 4.2. Leader Election and Group Reconfiguration

Once the leader fails, a new leader is elected through leader election. Since both group reconfiguration and event handling rely on group leader, leader election has the highest priority in the three routines. It interrupts any ongoing group reconfiguration or event delivery process. The leader election criterion is replica age, which increases by one after each group reconfiguration. Based on the assumption that node failure rate increase with time, the new leader is the youngest replica in the group.

Group reconfiguration adds new members for fault-tolerance. A group reconfiguration will be triggered once the group size is lower than $n$ and recover group size to $n + e$. Group reconfiguration has higher priority than event delivery, so that new members can be quickly added into a group. After group reconfiguration, the leader will also notify all senders of the new configuration.

In both leader election and group reconfiguration, the leader will decide the current states, namely $Q_d$, $E$, and $G$, and synchronize them to all replicas so that all replicas will load the same state after a leader election or a group reconfiguration, which is called state synchrony. For new replicas, the application state, the sequence of the last applied event $\lambda_c$, the time of the first cycle $t_0$, the start time $t_{start,s}$ for each sender $s$, and the sender set $S$ are also synchronized from the leader for initialization. The detailed algorithms of leader election and group reconfiguration are in Appendix B.

# 5. Virtual World Interaction

Virtual world interaction describes how users manipulate the state of virtual objects and perceive the state change in a shared simulated environment. Following the definition, a virtual world interaction includes two steps. First, a user modifies the state of an object through operations. Second, the new object state is synchronized to other interested users. This section extends the proposed event delivery approach for supporting interactions in Virtual Net.

## 5.1. Flow of Events and Updates

An object is replicated on multiple hosts (i.e., clients and mesh computers) if multiple users operate the same object. To facilitate object state consistency management in interaction, the copies of objects are classified to authoritative copies and non-authoritative copies. Each object can only have one authoritative copy but multiple non-authoritative copies. An authoritative copy is maintained by one mesh computer, e.g., the object owner's. The non-authoritative copies are maintained by the clients and the mesh computer of other interested users for fault-tolerance. Interest management has been intensively studied in [2], thus not discussed in this paper. It is only assumed that a user's interest scope is determined by her perception range in a virtual world, as illustrated in Figure 10.

By distinguishing authoritative copies from non-authoritative copies, object state management is simplified to managing the state of an authoritative copy and synchronizing the updated state from the authoritative copy to non-authoritative copies. For managing the authoritative copy, since the data is replicated to multiple nodes in a mesh computer, the fast event delivery approach is applied for maintaining the same state among these replicas.

From the perspective of an authoritative copy, an interaction includes receiving the event from one client, handling the event after it is delivered to the application, and multicasting the updated state to all interested hosts. Figure 9 illustrates the flow of events and updated states. The events of an object are only sent from the clients to the mesh computer which maintains the authoritative copy of the object, while updated states are broadcast by the mesh computers to all the non-authoritative copies. To support interaction, each mesh computer maintains two sets: the event sender set $S$ and the update recipient set $U$.

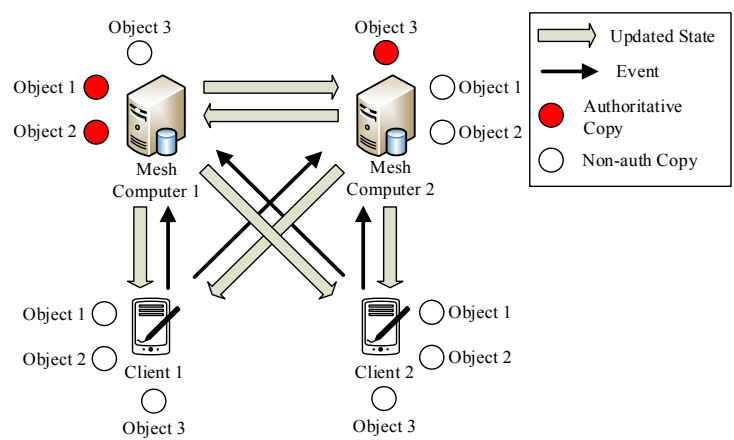

Figure 9: The flow of events and updated states in virtual world interaction.

## 5.2. Neighbors

To reduce overhead, each user only communicates with a limited number of peer users, called the neighbors. Due to user mobility, a user's neighbor may be frequently changed. A neighbor join happens when another user enters the perception range of a user. Likewise, a neighbor leave happens when a neighbor moves out of a user's perception range. For neighbour change, the key problem is to determine the same cycle of neighbor change on all replicas for their agreement on the cycle events. The join/leave cycle can be simply synchronized through consensus. However, high neighbor dynamics will increase the number consensus, resulting in high communication overhead and high interaction latency.

To apply the fast event delivery approach in neighbour change, the connectivity maintenance approaches of mutual notification [4] are employed. Specifically, two types of neighbour are introduced:

1.  Perception neighbor set ($N_p$): is the set of users and their virtual objects appearing in the perception range of a user.
2.  Connectivity neighbor set ($N_c$): is the set of users logically connected to the user.

Assume each user maintains a set of connectivity neighbors $N_c$. How to achieve it in a P2P virtual world can be found in [4]. A user (called *User i*) periodically exchanges its perception neighbor set $N_p$ with the connectivity neighbors. Once a connectivity neighbor finds that another user should / should not be in $N_p$, it will notify *User i*. To facilitate description, some abstract functions are introduced:

*   *Multicast*(*e, y, g*): Event *e* with sequence number *y* is sent to all replicas of group *g*.
*   *Handle*(*e*): Event *e*, which has been delivered to $Q_d$, is handled by the application.
*   *Time*(*t*): Set the timer to *t*, which will trigger a timeout event at *t*.
*   EVENT ← *c*: Assign content *c* to event EVENT.

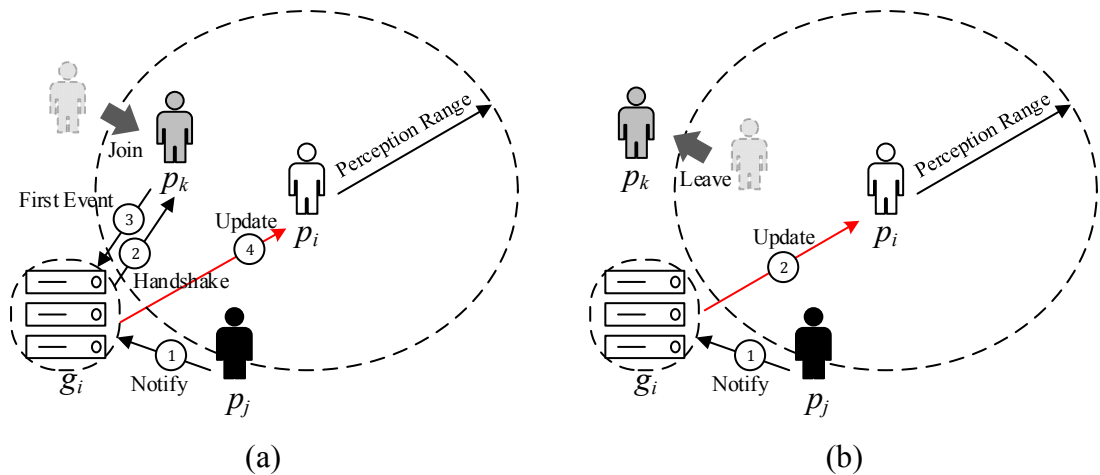

(a)          (b)

Figure 10: Neighbor change: (a) Neighbor Join; (b) Neighbor leave.

## 5.3. Neighbor Join

Suppose *User j* is one of the connectivity neighbor of *User i*. *User j* discovers that another user *User k* is in the perception range but not in the $N_p$ of *User i*, it will notify *User i* for adding the new neighbor with the following procedure. To distinguish clients from mesh computers, let $p_i$ be the client of *User i* and $r_i$ be the replica of group $g_i$ (i.e., the mesh computer of *User i*), the same for $p_j$ and $p_k$.

Step 1. $p_j$ *Multicast*(ADD_NEIGHBOR $\leftarrow$ ($p_k$, $G_k$), $y$, $g_i$).
Step 2. $r_i$ *Handle*(ADD_NEIGHBOR) for cycle $c$, $\forall r_i \in G_i$.
    a. $r_i$ modifies $S \leftarrow S \cup \{p_k\}$, $U \leftarrow U \cup \{p_k, G_k\}$.
    b. $r_i$ calculates $t_{start,k} = t_{recv,j}(y) + n \cdot \Delta t$ and $t_{recv,k}(1) = t_{start,k} + \Delta t$.
    c. $r_i$ calculates cycle $c_k = [(t_{recv,k}(1) - t_{now}) / \Delta t] + c$.
    d. $r_i$ *Time*($c_k$) for receiving the first event from $p_k$.
    e. $r_i$ sends (HANDSHAKE $\leftarrow$ ($t_{start,k}$, $G_i$)) to $p_k$.
Step 3. $p_k$ adds $G_i$ to the recipient list.
Step 4. $p_k$ Calculate $t_{send,k}(1) \leftarrow t_{start,k} - \Delta T_n^L$
Step 5. $p_k$ *Time*($t_{send,k}(1)$).
Step 6. $p_k$ *Multicast*($e$, $0$, $g_i$) at $t_{send,k}(1)$.
Step 7. $p_k$ *Time*($t_{start,k} + \Delta t$) for the next event.
Step 8. $r_i$ receives EVENT at $c_k$.
Step 9. $r_i$ delivers EVENT for $c_k$.

The neighbor join process is illustrated in Figure 10a. First, the connectivity neighbor $p_j$ sends the ADD_NEIGHBOR event to all replicas in $g_i$ for adding new neighbor $p_k$. Then, each replica $r_i$ modifies the event sender the and the update set recipient set, calculates the first event start time for event sending and receiving, and notify the new neighbor $p_k$. The client $p_k$ is timed to send the first event at $t_{recv,j}(y) + n \cdot \Delta t - \Delta T_n^L$ (where $n$, $\Delta t$, and $\Delta T_n^L$ are preconfigured). Meanwhile, the replicas of $g_i$ are waiting for the event of the first cycle $c_k$ at $t_{recv,k}$. Since the start time $t_{start,k}$ and the first event sequence number $j_0 = 0$ are known to both communication ends, they can individually calculate the time of subsequent event sending and receiving. At last, client $p_i$ learns the new neighbor through the update from $g_i$ and renders it to *User i*.

### 5.4. Neighbor Leave

The procedure of neighbor leave is similar to but simpler than the neighbor join procedure. Suppose *User j* is one of the connectivity neighbor of *User i*. When $c_i$ discovers that another user *User k* is out of the perception range but still in the $N_p$ of *User i*, it will notify *User i* with the following procedure.

Step 1. $p_j$ *Multicast*(RM_NEIGHBOR $\leftarrow$ ($k$), $y$, $g_i$).
Step 2. $r_i$ *Handle*(RM_NEIGHBOR) for cycle $c$, $\forall r_i \in G_i$.
    a. $r_i$ modifies $S \leftarrow S \setminus \{p_k\}$, $U \leftarrow U \setminus \{p_k, G_k\}$ for cycle $c + 1$.

The neighbor leave process is illustrated in Figure 10b. The connectivity neighbor $p_j$ sends the RM_NEIGHBOR event to all replicas in $g_i$ for removing the neighbor $p_k$, which can then remove $p_k$ and $G_k$ in both the event sender set and the event recipient set for cycle $c + 1$. Through the update from $g_i$, then client $p_i$ learns the leave of neighbor $p_k$ and removes $p_k$ from display.

# 6. Theoretical Verification

The correctness of the state update design is determined by the state of all replicas in a group, as well as the clients. Firstly, the correctness of leader election and group reconfiguration are verified, since they support the other propositions. The proof of all lemmas and theorems can be found in Appendix C.

**Lemma 6.1 (Leader Election Synchrony).** All the live replicas in $G$ maintain the same $Q_d$, $E$, and $G$ after leader election.

**Lemma 6.2 (Group Reconfiguration Synchrony).** All the live replicas in $G$ maintain the same $Q_d$, $E$, and $G$ after a group reconfiguration.

Next, without loss of generality, the correctness of the consensus protocol is verified for an arbitrary cycle $c$. The validity property and the integrity property [27] are not verified here, since they are not related to the main result and easy to be verified. Interested users can prove them. Here, only the agreement property is verified.

**Lemma 6.3 (Consensus Agreement).** If a live replica $r_i \in G$ delivers an event $e$ to $E(c)$ from a consensus instance for cycle $c$, then $e$ is eventually delivered to $E(c)$ by all the live replicas.

With the above lemmas, the main result can be obtained. But before it, an important property of the late event handling approach needs to be verified first.

**Lemma 6.4 ($\Omega(s, c)$ Agreement).** All the live replicas in $G$ expect to deliver the same set of events $\Omega(s, c)$ for sender $s \in S$ and cycle $c$.

Now, the main result of theoretical verification can be presented with the following theorem and corollary.

**Theorem 6.5 (Total-order Event Delivery).** If a live replica $r_i \in G$ delivers two different events $e_1$ and $e_2$ into $Q_d$ with $\lambda_1$ and $\lambda_2$, then $e_1$ and $e_2$ will eventually be delivered into $Q_d$ on all the live replicas with $\lambda_1$ and $\lambda_2$ being two non-negative integer numbers and $\lambda_1 \neq \lambda_2$.

**Corollary 6.6 (Replica Synchronization).** All the live replicas in $G$ maintains the same state of their virtual objects.

Another important result is the correctness of garbage collection, which is verified in Theorem 6.7.

**Theorem 6.7 (Garbage Collection Safety).** If event $e$ is removed from $Q_d$ on $r_i \in G$, then $e$ has been handled by the application on all the live replicas in $G$.

Based on Theorem 6.5, the correctness of the neighbor change procedures is shown with the following corollaries.

**Corollary 6.8 (Total-order Event Delivery with Sender Join).** All the live replicas in $G$ deliver the same first event $e_0$ from a neighbor $s$ with the same delivery sequence $\lambda_0$.

**Corollary 6.9 (Total-order Event Delivery with Sender Leave).** All the live replicas in $G$

deliver the same last event $e_\infty$ from a neighbor $s$ with the same delivery sequence $\lambda_\infty$.

# 7. Performance Analysis and Comparison

The performance of the proposed fast event delivery approach is studied in terms of synchronization delay and update loss rate. Three alternative approaches are introduced and compared with the proposed approach: the primary-backup approach, the reliable primary-backup approach, and the consensus-based total-order approach.

In the primary-backup approach [8], one replica is the primary replica and the rest are the backup replicas. The primary receives and handles all events, and then broadcasts updates to recipients. Meanwhile, the primary replica sends the received events to backups for fault-tolerance. In a reliable primary-backup, the primary broadcasts the update only after the events have been reliably synchronized to all backups. Note that the unreliable primary-backup approach does not ensure state consistency in case of primary failure. The consensus-based total-order approach [27] is similar to the proposed design, except that all events are delivered through consensus. Specifically, in each cycle, all replicas propose the received events within the cycle, the leader decides the events delivery order for the cycle.

Synchronization delay describes the time consumed in synchronizing the events over all live replicas. The primary-backup approach does not have a synchronization delay. In the reliable primary-backup approach, only 2 communication steps are involved in event synchronization: the primary broadcasts the events to all backups and collects the response from the backups. The consensus-based total-order approach needs one more communication step, as shown in Figure 8. In the proposed approach, Synchronization delay is factored by the probability $p_{sync}$ of triggering the consensus protocol.

Update loss rate describes the probability that a client does not receive the corresponding update after it sends an event to a mesh computer, due to event loss or update loss. In the primary-backup approach, update loss will occur as long as the channel between an event sender/update recipient and the primary replica is failed. In the consensus-based approach and the proposed approach, update loss occurs only when no replica receives the event or all replicas fail to send the update to a recipient. Moreover, assume late and out-of-order events are discarded in all the approaches.

The performance comparison of different approaches is shown in Table 2, in which $d_c$ denotes the delay in collecting a message from all replicas, $d_m$ denotes the reliable multicast delay, and $p_{loss}$ denotes the probability of message loss on a link.

Table 2 Performance Analysis and Comparison Results

| | Synchronization Delay | Update Loss Rate |
|---|---|---|
| Primary-backup | $0$ | $p_{loss} + (1 - p_{loss})p_{loss}$ |
| Reliable Primary-backup | $d_c + d_m$ | $p_{loss} + (1 - p_{loss})p_{loss}$ |
| Consensus-based | $d_c + 2d_m$ | $p_{loss}^n + (1 - p_{loss}^n)p_{loss}^n$ |
| Fast Event Delivery | $(d_c + 2d_m)p_{sync}$ | $p_{loss}^n + (1 - p_{loss}^n)p_{loss}^n$ |

The comparison result shows that, if $p_{sync}$ is small, i.e., transmission latency and clock offset are low, then the synchronization delay of the proposed approach is small and may even close to that of the unreliable primary-backup approach. Thus the proposed approach can opportunistically provide higher responsiveness than the consensus-based total-order approach and the reliable primary-backup approach.

For update loss rate, $p_{loss}^n + (1 - p_{loss}^n)p_{loss}^n < p_{loss} + (1 - p_{loss})p_{loss}$ for $n \geq 2$, which can be proved as follows.

First, let *n = 2*. Then,

$$p_{loss}^2 + (1 - p_{loss}^2)p_{loss}^2 - (p_{loss} + (1 - p_{loss})p_{loss})$$
$$= -p_{loss}(p_{loss} + 2)(p_{loss} - 1)^2$$
$$< 0.$$

Thus, $p_{loss}^n + (1 - p_{loss}^n)p_{loss}^n < p_{loss} + (1 - p_{loss})p_{loss}$ for *n = 2*.

Second, let $f(x) = p_{loss}^x + (1 - p_{loss}^x)p_{loss}^x$, $x \in \mathbb{R}$. Then,

$$\frac{\partial}{\partial x}(p_{loss}^x + (1 - p_{loss}^x)p_{loss}^x)$$
$$= p_{loss}^x ln(p_{loss}) + (1 - p_{loss})^x p_{loss}^x ln(1 - p_{loss})$$
$$+ (1 - p_{loss})^x p_{loss}^x ln(p_{loss})$$
$$< 0.$$

Thus, $p_{loss}^n + (1 - p_{loss}^n)p_{loss}^n$ monotonically decreases along with *n*.

Therefore, $p_{loss}^n + (1 - p_{loss}^n)p_{loss}^n < p_{loss} + (1 - p_{loss})p_{loss}$ for $n \geq 2$. This shows that the consensus-based total-order approach and the proposed approach have lower update loss rate than the primary-back approaches.

# 8. Experiments and Results

## 8.1. Simulation Setup

The proposed model is evaluated by simulating a distributed computing. Experiments are run in OMNeT++ to simulate message transmission in a network and event-based programming[3]. The simulation is run by sending events from 10 clients to a replica group, representing 10 neighbors. The replica group size is configured to 5. Cycle length is set to 200ms. In experiments, each client sends more than 9000 events to the replica group, which can simulate a half-hour game session with 200ms user operation inter-arrival time, applicable to most game genres [30]. After events are sorted and handled, updates will be transmitted to clients for collecting the statistic result. New replicas are generated by the Rendezvous, if the group size is lower than the availability threshold new replicas will be generated by the Rendezvous.

The network traffic model includes packet latency and packet loss. The packet loss rate is varied to simulate different rate of message loss rate $p_{loss}$. The packet delay is calculated by one-trip communication delay and network jitter. To facilitate simulation, network traffic is generated by a generating function from the analytical result of real data. [31] suggests that the one-way delay between two hosts $H_1$ and $H_2$ can be modelled by *delay(H₁, H₂) = D_{min} + jitter*, where $D_{min}$ is the minimum single-trip delay. *jitter* is the network jitter caused by network congestion. In the experiments, $D_{min}$ is configured to 50ms and *jitter* is modelled by a truncated normal distribution and varied to simulate the variation of network latency.

---

[3] Simulation code at https://github.com/sunniel/VirtualNetEventHandling.

To simulate replica failure, replica dynamics is characterized by session length which measures the length of time that a peer is continuously connected to a given P2P network, from its arrival to its departure [32]. Session length of P2P applications can be depicted by different stochastic models. [32] shows that Weibull distribution or log-normal distribution fits the observation best. In this study, replica session length is modelled by a Weibull distribution with the mean value of half hour.

## 8.2. Experiment Results for Overall Performance

To verify the overall performance of event handling, three alternative event delivery approaches are implemented, including primary-backup, consensus-based total-order, and fast event delivery, for comparing their performances. Reliable primary-backup is not included in the comparison, since its performance is in-between the unreliable primary-backup approach and the consensus-based approach.

First, responsiveness is evaluated by comparing the interaction latency in different approaches. Interaction latency includes both the round-trip end-to-end delay between an event sender and a group of replicas and the synchronization delay. The mean value of network jitter is fixed to 50ms, and the standard deviation is changed from 50ms to 250ms to simulate the scenario that events occasionally come late and out-of-order. The experiment result in Figure 11 shows that the fast event delivery approach provides much lower latency than the consensus-based total-order approach. Especially when the network latency is small, the responsiveness of the proposed approach is close to the primary-backup model. This is because the rate of triggering the consensus protocol decreases, when most events arrive before the end of cycles.

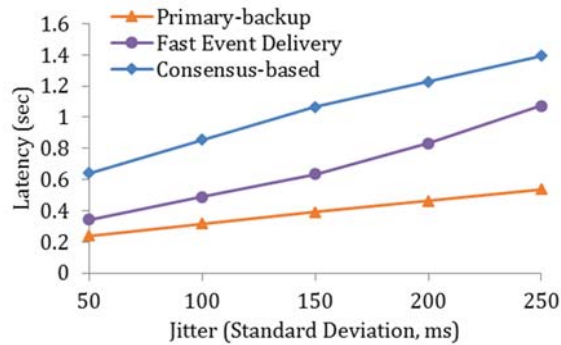

Figure 11: Interaction latency comparison in different models.

Second, end-to-end update delivery rate is evaluated by varying the message drop rate to simulate the change of $p_{loss}$ from 0.3 to 0.7. In an asynchronous network, message loss cannot be distinguished from long message delay. Thus, update delivery timeout is used to cover both situations, which is configured to 5 seconds. The mean value of network jitter is fixed to 50ms to eliminate the interference of late events.

Figure 12 shows the update delivery rate of the three different approaches. Specifically, the update delivery rate of the primary-backup approach is much lower than the other two approaches. Moreover, it drops quickly from around 0.7 to below 0.3 with the increase of $p_{loss}$, showing the rapid increase of update loss rate. In contrast, the update delivery rate of the consensus-based total-order approach and the proposed approach (overlapped) can remain high. Before $p_{loss}$ is lower than 0.5, the update delivery rate of these two approaches is close to 1. When $p_{loss}$ is lower than 0.5, the drop of the update delivery rate of them becomes evident. This is because, with the increase of message drop rate, more messages are received by none of the replicas.

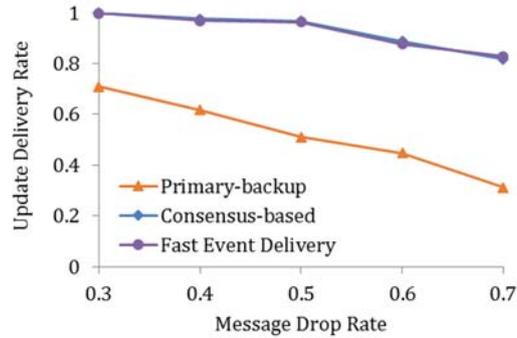

Figure 12: Update delivery rate with different message drop rate.

In the same experiment, interaction latency is also studied with the change message drop rate. Figure 13 shows that, with the increase of message drop, a replica has a higher chance to miss the cycle events, such that $\Pi(s, c) \neq \Omega(s, c)$ and more consensus instances are triggered by replicas for event synchronization. This means that increasing message drop has a similar effect of increasing network jitter on the fast event delivery approach in interaction latency.

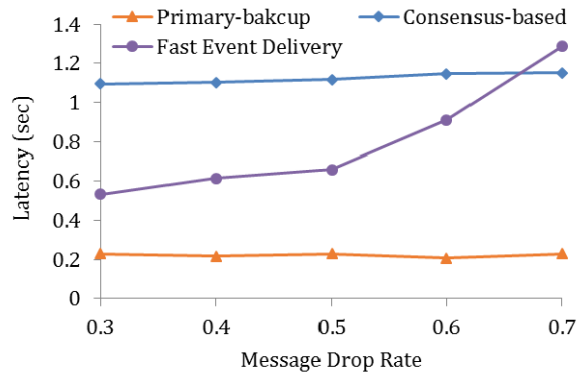

Figure 13: Interaction latency with different message drop rate.

### 8.3. Experiment Results for Individual Improvements

#### 8.3.1. Performance of Late Event Handling

The experiment result in this section verifies the late event handling approach. The proposed approach is compared with the simple discard approach which simply discards any late events. In the experiment, the timing of event sending is modified with clock error which is modelled with a normal distribution ($\mu = 0$). The standard deviation of the clock error is changed from 0 to 400ms to increase the rate of late event. Moreover, network jitter is fixed to 50ms and the message drop rate $p_{loss}$ is fixed to 0 to eliminate their interference to the result.

Figure 14 shows that the proposed approach has a much higher update delivery rate than the simple discard approach. Especially when the clock error is higher than 300ms, simply discarding late events results in that almost no message is delivered, since most events will come late. On the other hand, through the proposed approach, the update delivery rate can be maintained as high as close to 1. But the update delivery rate is lower than 1, because a few late events do not meet the deliverability condition.

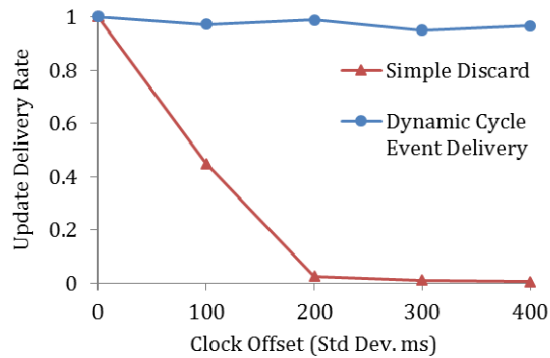

Figure 14: Update delivery rate with different late event handling approaches.

### 8.3.2. Performance of Garbage Collection

Garbage collection is tested to verify its effectiveness. The main purpose of the experiment is to show that the proposed mechanism can effectively limit the length of $Q_d$ from over-growth. Thus, the experiment is conducted with two different settings: one with garbage collection and the other without garbage collection. The increase of the delivery queue length ($Q_d$) is observed and compared for the two settings. Network jitter is fixed to 50ms and the message drop rate $p_{loss}$ is configured to 0 to remove the interference of event loss in both settings, so that the length of $Q_d$ is only determined by the number of events and garbage collection. In the second setting, the length of the garbage collection cycle is fixed to 5 seconds. The experiment result is shown in Figure 15. In the case that garbage collection is not applied, the length of $Q_d$ quickly increases from several thousand events to tens of thousands of events within 500 seconds. On the other hand, if garbage collection is applied, the change of $Q_d$ is restrained within 300 hundred events. The comparison result shows that the proposed garbage collection protocol can effectively prevent $Q_d$ from unlimited growth or even overflow.

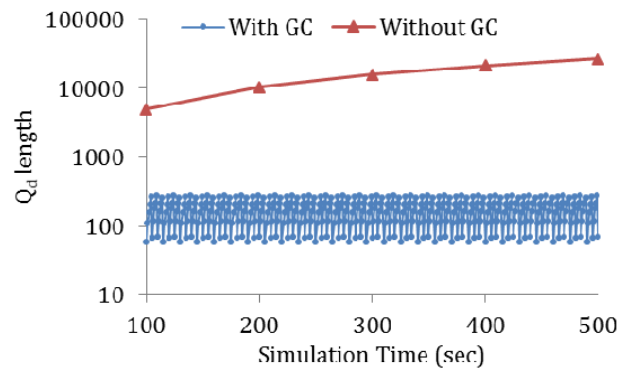

Figure 15: Delivery queue ($Q_d$) length along with simulation time with and without garbage collection (GC).

Moreover, the cycle length of the gossip protocol is changed to show the control of the protocol on $Q_d$ length. The experiment result is shown in Figure 16. When the cycle length of the gossip protocol increases from 1 second to 10 seconds, the length of $Q_d$ changes from around 50 events to around 500 events. This experiment result shows that the length of $Q_d$ is approximately linear to the cycle length of the gossip protocol. It implies that the length of $Q_d$ can be effectively controlled by changing the cycle length.

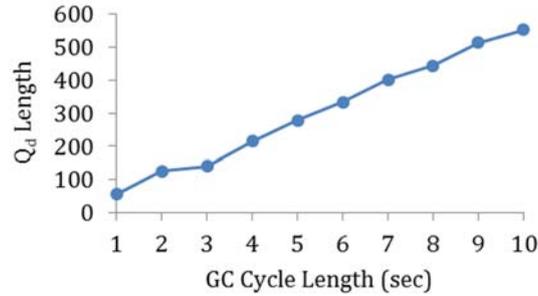

Figure 16: Trend of delivery queue ($Q_d$) length with different garbage collection (GC) cycle length.

The length of $Q_d$ affects the time of state synchronization in leader election and group reconfiguration, because all the data in $Q_d$ need to be transmitted to new replicas to recover their states. Since event handling is suspended by leader election and group reconfiguration, longer $Q_d$ will increase the time of data transmission, thus increases the interaction latency experienced by users.

An experiment was conducted to verify the relation of garbage collection and interaction latency by changing the gossip cycle length from 2 seconds to 10 seconds. The life span of replicas are also shortened to increase the number of leader election and group reconfiguration. The result in Figure 17 shows that, though most latencies are low, there are some latency bursts which are high. These bursts correspond to the latencies mainly generated from leader election and reconfiguration. It can be found that the bursts becomes high along with the garbage collection cycle, for more events to be transmitted from $Q_d$. Moreover, longer duration of state synchronization not only increases interaction delay, but also brings higher risk of group failure (Figure 17c), more replicas could be failed during this period. Thus, a proper garbage collection cycle needs be selected related to the replicas' statistical states, such as their mean time-to-failure. We leave it for our future work.

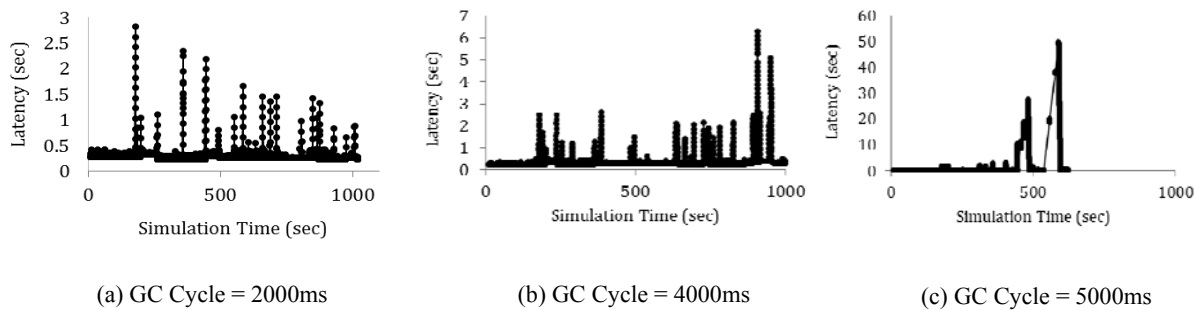

(a) GC Cycle = 2000ms        (b) GC Cycle = 4000ms        (c) GC Cycle = 5000ms

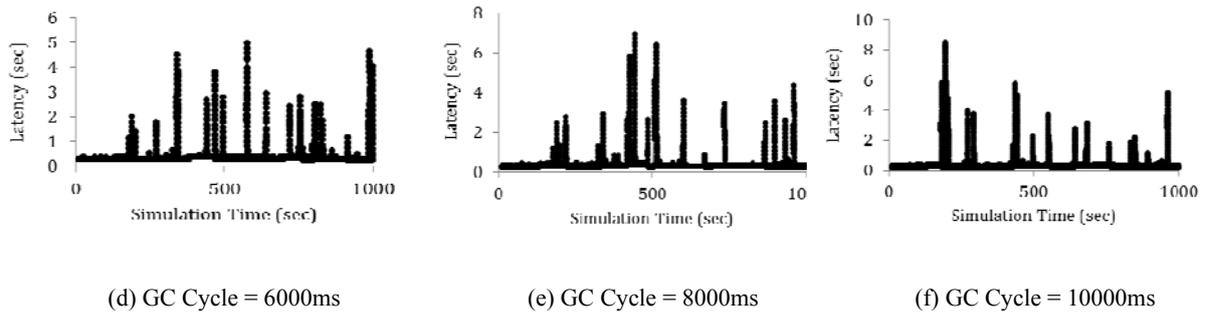

(d) GC Cycle = 6000ms      (e) GC Cycle = 8000ms      (f) GC Cycle = 10000ms

Figure 17 Interaction latency affected by garbage collection (GC) with different cycle lengths.

### 8.3.3. Performance of Time Synchronization

The performance improvement through time synchronization is shown in Figure 18 and **Error! Reference source not found.**. The main purpose of the experiment is to show that the time synchronization mechanism has positive effect on interaction latency reduction and garbage collection. Thus, the experiment compares the performance of event handling in two different settings: one with time synchronization and the other without time synchronization. Clock error is added into the timing of event sending to simulate the scale of clock synchronization loss between the event senders and recipients. Network jitter is fixed to 50ms and the message drop rate $p_{loss}$ is fixed to 0 to eliminate their interference to the result.

Figure 18 shows that if the clocks between event senders and event recipients are out-of-sync, the increase of interaction latency along with clock offset, because time synchronization loss increases the chance of triggering consensus for event delivery. Note that the interaction latency increase with clock offset is almost linear, clearly showing the impact of synchronization loss. In contrast, if time synchronization is applied before event transmission, the interaction latency is less than 1 second. This is because the number consensus can be reduced and thus the communication steps for event delivery can be minimized accordingly.

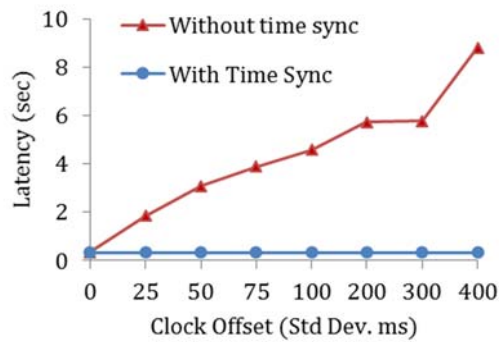

Figure 18: Interaction latency with different clock offset.

### 8.4. Discussion

The experiment results for the overall performance show similar results to the theoretical analysis. Specifically, the proposed fast event delivery approach is reliable and can provide opportunistically high responsiveness, compared with the consensus-based total-order approach and the primary-backup approach, because this approach has the highest update delivery rate, and almost the same interaction latency as the primary-backup approach when the network latency is low. Thus, the overall performance of the proposed approach is better than the other two alternative approaches. The results also imply that, in practice, it is better to select cache nodes which are close to clients so that most cycle events can arrive on time at all replicas and they can be delivered without consensus. Then, interaction latency can be minimized.

The experiment results for the individual improvements show their effectiveness on system performance improvement. Specifically, the experiment result of late event handling shows that the proposed dynamic cycle event delivery approach can largely increase update delivery rate. High update delivery rate can reduce the chance of event re-sending which will lower down responsiveness in interaction and impact user experience. The experiment result of garbage collection shows its effectiveness in limiting the length of the event delivery queue. This is important, because it can not only reduce the scale of latency burst generated by state synchronization but also decrease the chance of group failure. Lastly, the evaluation result of time synchronization shows its importance. Without the time synchronization between event senders and event recipients, the advantage of the fast event handling approach diminishes.

## 9. Conclusions

With the popularity of mobile virtual worlds, scalability becomes an outstanding challenge in infrastructure development. The possibility of P2P technology is discussed to address the scalability problem. Different from existing P2P virtual worlds, client unreliability raises a new problem in mobile settings. This paper tries to solve the problem with a new hierarchical P2P computing model. Yet, rather than introducing every detail of the computing model, we focus on object state update to avoid reinventing wheel. The core problem of state update is to maintain the replica state consistency without compromising system responsiveness. To address the problem, a fast event delivery approach is proposed. Based on this approach, we introduce the new virtual world interaction model to enable the interaction between multiple users.

Our work is important in providing a scalable infrastructure for mobile P2P virtual worlds. Based on the proposed Virtual Net architecture, there are some new research problems for building virtual world applications. First, our current approach still has the limitation in high responsiveness, since it belongs to the opportunistic category. To further improve system responsiveness without compromising state consistency, we plan to employ conflict-free replicated data types (CRDT) [33] to replace the consensus approach in event handling. With CRDT, events can be delivered in any sequence. However, events delivery in different sequences may cause user confusion with respect to continues events, such as avatar movement. Thus, it can be expected that the problem will be a combination of human-computer interaction (HCI) distributed computing. Moreover, the future study also includes the application and adaptation of cloud-fog computing techniques for contributed resource management, including cache node allocation, and P2P virtual world techniques to provide a complete and practical mobile P2P virtual world solution.

## Conflicts of Interest

The author(s) declare(s) that there is no conflict of interest regarding the publication of this paper.

# Acknowledges

This research is partially supported by the University of Macau Research Grant No. MYRG2017-00091-FST and MYRG2015-00043-FST.

# References

[1] B. Knutsson, H. Lu, W. Xu, and B. Hopkins. (2004). "Peer-to-peer support for massively multiplayer games," in IEEE INFOCOM 2004, Hong Kong, pp. 107. doi: 10.1109/INFCOM.2004.1354485.

[2] T. Malherbe. (2016). "A Comparative study of interest management schemes in peer-to-peer massively multiuser networked virtual environment," MEng. Thesis, Stellenbosch University, Available: http://hdl.handle.net/10019.1/100061.

[3] A. Yahyavi and B. Kemme. (2013). "Peer-to-peer architectures for massively multiplayer online games: A survey," ACM Computing Surveys (CSUR), 46(1), pp. 9.

[4] E. Buyukkaya, M. Abdallah, and G. Simon. (2015). "A survey of peer-to-peer overlay approaches for networked virtual environments," Peer-to-peer networking and applications, 8(2), pp. 276-300.

[5] S. Y. Hu, J. F. Chen, and T. H. Chen, (2006). "VON: a scalable peer-to-peer network for virtual environments," IEEE Network, 20(4), pp. 22-31.

[6] L. Ricci, L. Genovali, E. Carlini, and M. Coppola. (2015). "AOI-cast in distributed virtual environments: an approach based on delay tolerant reverse compass routing. Concurrency and Computation: Practice and Experience," 27(9), pp. 2329-2350.

[7] A. Yahyavi, K. Huguenin, J. Gascon-Samson, J. Kienzle, and B. Kemme. (2013, Jul.). "Watchmen: Scalable cheat-resistant support for distributed multi-player online games," In IEEE 33rd International Conference on Distributed Computing Systems (ICDCS), Philadelphia, PA, pp. 1-10. doi: 10.1109/ICDCS.2013.62

[8] H. A. Engelbrecht, J. S. Gilmore. (2017). "Pithos: Distributed storage for massive multi-user virtual environments," ACM Transactions on Multimedia Computing, Communications, and Applications (TOMM), 13(3), pp. 31.

[9] P. E. Ross. (2009). "Cloud computing's killer app: Gaming," IEEE Spectrum, 46(3), pp. 14-14.

[10] W. Cai, R. Shea, C. Y. Huang, K. T. Chen, J. Liu, V. C. M. Leung, and C. H. Hsu. (2016, Aug). "A Survey on Cloud Gaming: Future of Computer Games," IEEE Access, 4, pp. 7605-7620.

[11] S. Choy, B. Wong, G. Simon, and C. Rosenberg. (2012, November). "The brewing storm in cloud gaming: A measurement study on cloud to end-user latency," In Proceedings of the 11th Annual Workshop on Network and Systems Support for Games (NetGames'12). IEEE, Piscataway, NJ, USA, Article 2, pp. 1-6.

[12] L. M. Vaquero and L. Rodero-Merino. (2014). "Finding your way in the fog: Towards a comprehensive definition of fog computing." ACM SIGCOMM Computer Communication Review, 44(5), pp. 27-32.

[13] M. R. Anawar, S. Wang, M. A. Zia, A. K. Jadoon, U. Akram, and S. Raza. (2018). "Fog Computing: An Overview of Big IoT Data Analytics," Wireless Communications and Mobile Computing, vol. 2018, Article ID 7157192, pp. 1-22. doi: 10.1155/2018/7157192.

[14] H. T. Dinh, C. Lee, D. Niyato, and P. Wang. (2013). "A survey of mobile cloud computing: architecture, applications, and approaches," Wireless communications and mobile computing, 13(18), pp. 1587-1611.

[15] E. C. P. Neto, G. Callou, F. Aires. (2017). "An algorithm to optimise the load distribution of fog environments," In IEEE International Conference on Systems, Man, and Cybernetics (SMC), Banff, AB, pp. 1292-1297. doi: 10.1109/SMC.2017.8122791.

[16] Y. Lin and H. Shen. (2017). "CloudFog: Leveraging fog to extend cloud gaming for thin-client MMOG with high quality of service," IEEE Transactions on Parallel & Distributed Systems, 28(2), pp. 431-445.

[17] N. Wang, B. Varghese, M. Matthaiou, and D. S. Nikolopoulos. (2017). "ENORM: A framework for edge node resource management," IEEE Transactions on Services Computing, in press, doi: 10.1109/TSC.2017.2753775.

[18] M. Claypool and K. Claypool. (2006). "Latency and player actions in online games," Communications of the ACM, 49(11), pp. 40-45.

[19] E. K. Lua, J. Crowcroft, M. Pias, R. Sharma, and S, Lim. (2005). "A survey and comparison of peer-to-peer overlay network schemes," IEEE Communications Surveys & Tutorials, 7(2), pp. 72-93.

[20] E. Carlini, L. Ricci, and M. Coppola. (2013). "Flexible load distribution for hybrid distributed virtual environments," Future Generation Computer Systems, 29(6), pp. 1561-1572.

[21] R. Bhagwan, K. Tati, Y. Cheng, S. Savage, and G. M. Voelker. (2004, March). "Total Recall: System Support for Automated Availability Management," In Proc. 1st Conf. Networked Systems Design and Implementation (NSDI).

[22] R. C. Merkle. (1988). "A digital signature based on a conventional encryption function," In Pomerance C. (eds) Advances in Cryptology — CRYPTO'87. CRYPTO 1987. Lecture Notes in Computer Science, vol 293. Springer, Berlin, Heidelberg. doi: 10.1007/3-540-48184-2_32

[23] C. C. Erway, A. Küpçü, C. Papamanthou, R. Tamassia. (2015). "Dynamic provable data possession," ACM Transactions on Information and System Security (TISSEC), 17(4), pp. 15.

[24] C. Symborski, (2008). "Scalable user content distribution for massively multiplayer online worlds," Computer, 41(9), pp. 38-44.

[25] B. Shen, J. Guo, and L. X. Li. (2017). "Cost optimization in persistent virtual world design," Information Technology and Management, 19(3), pp. 1-15.

[26] F. B. Schneider, (1990). "Implementing fault-tolerant services using the state machine approach: A tutorial," ACM Computing Surveys (CSUR), 22(4), pp. 299-319.

[27] C. Cachin, R. Guerraoui, and L. Rodrigues. (2011). "Introduction to reliable and secure distributed programming (2nd ed.)," Springer Publishing Company, Incorporated.

[28] A. Chandler and J. Finney. (2005, June). "Rendezvous: supporting real-time collaborative mobile gaming in high latency environments," In Proceedings of the 2005 ACM SIGCHI International Conference on Advances in computer entertainment technology (ACE'05). ACM, New York, NY, USA, pp. 310-313. doi: 10.1145/1178477.1178534.

[29] D. L. Mills. (1991). "Internet time synchronization: the network time protocol," IEEE Transactions on communications, 39(10), pp. 1482-1493.

[30] X. Che and B. Ip. (2012). "Packet-level traffic analysis of online games from the genre characteristics perspective," Journal of Network and Computer Applications, 35(1), pp. 240-252.

[31] S. Kaune, K. Pussep, C. Leng, et al. (2009, February). "Modelling the internet delay space based on geographical locations," In 17th Euromicro International Conference on Parallel, Distributed and Network-based Processing, Weimar, 2009, pp. 301-310. doi: 10.1109/PDP.2009.44.

[32] D. Stutzbach and R. Rejaie. (2006). "Understanding churn in peer-to-peer networks," In Proceedings of the 6th ACM SIGCOMM conference on Internet measurement (IMC'06). ACM, New York, NY, USA, pp. 189-202. doi: 10.1145/1177080.1177105.

[33] M. Shapiro, N. Preguica, C. Baquero, and M. Zawirski. (2011). "Conflict-free Replicated Data Types," [Research Report] RR-7687, INRIA. 2011, pp.18. Available: https://hal.inria.fr/inria-00609399v2.

## Appendix A Fast Event Delivery Protocols

The full set of the fast event delivery protocols are described in this appendix, which includes event collection, event delivery, and consensus. The payload of an event could be the operation of the event sender, *Empty*, or ⊥ (called ⊥ event). Note that if a reliable communication channel is required in a function, the keyword *Reliably* will be added before the send or broadcast operation. The implementation of a reliable channel can be found in [27]. Message names are capitalized and messages could contain some parameters. The notations used in the pseudocodes are listed in Table 3. Particularly, $G \cap R$ denotes the set of live replicas.

Table 3 Notations in Protocols

| Notations | Descriptions |
|---|---|
| $r_i$ | Replica $i$ |
| $r_L$ | Group leader |
| $r_c$ | Group leader candidate |
| $G$ | The set of group members |
| $R$ | The set of live replicas |
| $S$ | The set of senders |
| *epoch* | The epoch number based on leader change |
| *cid* | Configuration ID |
| *age* | Age of a replica |
| $c$ | Cycle number $c$ |
| $\Delta t$ | Cycle length |
| $E$ / $E(c)$ | The decided events / Decided events of $c$ |

| $Q_d$ | Delivery queue, containing the sorted events to be delivered to the application in sequence |
|---|---|
| $Q_r$ | The received events |
| $Q_p$ | The temporary buffer for event collection |
| $s$ | Sender $s \in S$ |
| $Index(s)$ | The index of $s$ in $S$ sorted by sender ID |
| $Seq(e)$ | The sequence number of event $e$ |
| $Sender(e)$ | The sender of event $e$ |
| $e$ | Event e, in format of ($Sender(e)$, $Seq(e)$, $Payload(e)$) where $Payload$(e) denotes the payload of the event |
| $Seq(s, c)$ | The event sequence number of c from $s$ |
| $e(s, j)$ | The event $e$ of sequence number $j$ from $s$. |
| $MinSeq(s, c)$ | The lowest sequence number of all undelivered events from $s$ in $c$ |
| $MaxSeq(s, c)$ | The sequence number of the last delivered non-empty event from $s$ in $c$ |
| $Consensus(c)$ | Return *true* if there is a consensus on the fly for $c$ |
| $P$ | The set of cycles waiting consensus |
| $Z$ | The set of cycles in consensus |
| $LE$ | The flag of leader election |
| $GE$ | The flag of group reconfiguration |

In each cycle $\Delta t$, the event collection protocol (Algorithm 4) periodically collects the received events from all senders in $S$ in the receiving buffer $Q_r$ and to a temporary buffer $Q_p$. As described in the Late Event Handling section, not only the cycle events (i.e., $Seq(e) = Seq(i, c)$, Line 8-11) but also the late and still deliverable events are collected (Line 12-15). If any expected event is not received, a ⊥ event will be assigned to the corresponding event sequence number (Line 10). Likewise, a late event replaces an empty event with the same sequence number in $Q_p$ (Line 14-15). All undeliverable events will be discarded in event receiving (Line 3).

| Algorithm 4. Event Collection |
|---|
| 1.  <u>On replica $r_i$:</u> |
| 2.  Upon EVENT $e$ |
| 3.      If $Seq(e) > MaxSeq(Sender(e), c - 1)$, then |
| 4.          $Q_r \leftarrow Q_r \cup \{e\}$ |
| 5. |
| 6.  Upon cycle TIMEOUT |
| 7.      $c \leftarrow c + 1$ |
| 8.      For each $s \in S$, |
| 9.          If $e(s, Seq(s, c)) \notin Q_r$, then |
| 10.             $e \leftarrow (s, Seq(s, c), \perp)$ |
| 11.             $Q_p \leftarrow Q_p \cup \{(c, e)\}$ |
| 12.         For each $c' \in [0, c] \wedge s \in S \wedge e(s, c') \notin Q_d$,        // Collect late and deliverable events |
| 13.             If $e(s, Seq(s, c')) \in Q_r$ |
| 14.                 $Q_p \leftarrow Q_p \setminus \{e \mid e = (s, Seq(s, c'), \perp)\}$ |
| 15.                 $Q_p \leftarrow Q_p \cup \{(c, e(s, Seq(s, c')))\}$ |
| 16.                 $Q_r \leftarrow Q_r \setminus \{(e) \mid e(s, \text{Seq}(s, c)) \in Q_p\}$ |
| 17.     Reset Timer cycle $\leftarrow \Delta t$ |

The event delivery protocol (Algorithm 5) is executed by detecting the condition satisfaction of a cycle. If the events of cycle $c - 1$ has been delivered and events of cycle c has either 1) been collected or 2) decided from a consensus instance (Line 2), then the cycle $c$ satisfies the condition of triggering the event delivery protocol. Thus, the execution of event delivery is asynchronous to event collection. For distributed agreement, the protocol firstly checks the second condition to ensure that the consensus result will be applied on all replica. If the cycle is decided by a consensus instance, the decided events will be delivered not matter there is any non-empty event newly received for the cycle (Line 3-4). Otherwise, events will be delivered from $Q_p$. If all expected events have been collected (Line 6-8, 13-14), then they will be delivered to $Q_d$ in the sequence of $\gamma$ for cycle $c$. The range [$MinSeq(s, c)$, $Seq(s, c)$] specifies the deliverable sequence number for each sender s and cycle c. The calculation of $\gamma$ (Line 16) which ensures that all replicas can deliver the concurrent events from different senders in the same sequence. If there is any $\perp$ event, a query message will be sent to the group leader. The set union $\bigcup_{s \in S} MinSeq(s, c)$ and $\bigcup_{s \in S} Seq(i, c)$ specify the lowest deliverable sequence number and the sequence number of the cycle respectively for all senders.

The leader checks locally the receipt of the events for the requested cycle in $Q_p$ and $Q_d$. If all the expected events (for each sender s and each expected sequence number [$MinSeq(s, c)$, $Seq(s, c)$]) of the requested cycle have been received, it will reply them the requesting replica (Line 22-27). Otherwise, the leader will initialize a new consensus instance for the cycle (Line 28-29).

| Algorithm 5. Event Delivery |
| --- |

| | |
| --- | --- |
| 1. | On replica $r_i$: |
| 2. | Upon $(Q_d(c - 1) \neq \emptyset) \wedge (E(c) \neq \emptyset \vee Q_p(c) \neq \emptyset) \wedge \neg Consensus(c)$ |
| 3. | If $E(c) \neq \emptyset$, then |
| 4. | $D \leftarrow E(c)$ |
| 5. | Else |
| 6. | For each $s \in S \wedge j \in [MinSeq(s, c), Seq(s, c)]$, |
| 7. | $c' \leftarrow c - (Seq(s, c) - j)$ |
| 8. | $T \leftarrow T \cup \{(c, e(s, c')) \mid (c', e(s, j)) \in Q_p\}$ |
| 9. | If $\{(r, e) \mid Payload(e) = \perp \} \subseteq T$, then |
| 10. | QUERY $\leftarrow (c, \bigcup_{s \in S} MinSeq(s, c), \bigcup_{s \in S} Seq(s, c))$ |
| 11. | Reliably send QUERY to $r_L$ |
| 12. | End the procedure |
| 13. | Else |
| 14. | $D \leftarrow D \cup T$ |
| 15. | For each $(c, e)$ in $D$, |
| 16. | $\gamma \leftarrow Seq(e) + \sum_{k=1}^{Index(Sender(e))-1} (Seq(k, c) + 1)$ |
| 17. | $Q_d \leftarrow Q_d \cup \{(c, \gamma, e)\}$ |
| 18. | $c \leftarrow c + 1$ |
| 19. | |
| 20. | On leader $r_L$: |
| 21. | Upon QUERY$(c, \bigcup_{s \in S} MinSeq(s, c), \bigcup_{s \in S} Seq(s, c))$ from $r_i$ |
| 22. | For each $s \in S \wedge j \in [MinSeq(s, c), Seq(s, c)]$, |
| 23. | $c' \leftarrow c - (Seq(s, c) - j)$ |
| 24. | $R \leftarrow R \cup \{(c, e) \mid (c', e(i, j)) \in Q_p \vee (c, \gamma, e(s, j)) \in Q_d\}$ |
| 25. | If $\{(c, e) \mid Payload(e) = \perp \} \subseteq R$, then |
| 26. | QUERY_REPLY $\leftarrow (c, R)$ |
| 27. | Reliably send QUERY_REPLY to $r_i$ |
| 28. | Else |
| 29. | $P \leftarrow P \cup \{(c, \bigcup_{s \in S} MinSeq(s, c), \bigcup_{s \in S} Seq(s, c))\}$ |

The consensus protocol is run instantiated for each requested cycle c. Note that the consensus protocol is only executed when there is not leader election (LE) or group reconfiguration (GE). Also, a message from a previous leader or a previous group configuration is not be processed for consistency. Thus, these preconditions are added in all message handling procedures in the consensus protocol (Line 8, 11, 19, 26). First, the flags *LE* and *GE* are checked to ensure that the previous leader election and group reconfiguration have been finished. Then, the message sender's epoch and configuration ID are compared with the local epoch and configuration ID via adding the sender epoch and configuration ID to each message.

The consensus protocol is described in the Total-order Event Delivery section in detail. A replica replies to the leader for the query of the events for cycle $c$ only when the replica has passed the event collection of cycle $c$, which requires: 1) The decided events for cycle $c - 1$ has been delivered, if there is any; 2) The events collection for cycle $c$ has been done. The leader will decide the events for $c$ only after all proposals are received from all live replicas (Line 11). The Decide function determined the events of a given cycle for each event sender (Line 29-33) by the given set of received events for $c$ from all replicas. For each sender $s$ and event sequence number $j$, if all replicas propose $\perp$, then the payload of the event $e(s, j)$ will be decided with *Empty* (Line 33-34). Otherwise, the event payload will be decided with the value of the proposal from any replica (Line 30-31).

| Algorithm 6. Consensus |
| --- |

1. <u>On leader $r_L$:</u>
2. Upon $P \neq \emptyset \land CR = false \land LE = false$
3.     For each $(c, \bigcup_{s \in S} MinSeq(s, c), \bigcup_{s \in S} Seq(s, c)) \in P \land c \notin Z$,
4.         $Z \leftarrow Z \cup \{c\}$
5.         QUERY $\leftarrow$ $(epoch, cid, c, \bigcup_{s \in S} MinSeq(s, c), \bigcup_{s \in S} Seq(s, c))$
6.         Reliably send QUERY to $G \cap R$
7.
8. Upon QUERY_RESULT$(epoch', cid', r, W_i)$ from $r_i \land epoch' = epoch \land cid' = cid \land CR = false \land LE = false$
9.     $Q \leftarrow Q \cup \{(r_i, c, W_i)\}$
10.
11. Upon $G \cap R \subseteq \{r_i \mid (n_i, c, W) \in Q\} \land CR = false \land LE = false$
12.     $R' \leftarrow Decide(c, Q)$
13.     $P \leftarrow P \setminus \{(c, \bigcup_{s \in S} MinSeq(s, c), \bigcup_{s \in S} Seq(s, c))\}$
14.     $Z \leftarrow Z \setminus \{c\}$
15.     DECISION $\leftarrow$ $(epoch, cid, c, W')$
16.     Reliably broadcast DECISION to $G \cap R$
17.
18. <u>On replica $r_i$:</u>
19. Upon QUERY$(epoch', cid', r, \bigcup_{s \in S} MinSeq(s, c), \bigcup_{s \in S} Seq(s, c))$ from $r_L \land epoch' = epoch \land cid' = cid \land GR = false \land LE = false$
20.     For each $s \in S \land j \in [MinSeq(s, c), Seq(s, c)]$,
21.         $c' = c - (Seq(s, c) - j)$
22.         $W \leftarrow W \cup \{(c, e(s, j)) \mid (c', e(s, j)) \in Q_p \lor (c, \gamma, e(s, j)) \in Q_d \}$
23.     QUERY_RESULT $\leftarrow$ $(epoch, cid, c, W)$
24.     Reliably send QUERY_RESULT to $r_L$
25.
26. Upon DECISION$(epoch', cid', c, W')$ from $r_L \land epoch' = epoch \land cid' = cid \land GR = false \land LE = false$
27.     $E \leftarrow E \cup W'$
28.
29. Decide$(c, Q)$
30.     For each $s \in S$ and $j \in [MinSeq(s, c), Seq(s, c)]$,
31.         If $\exists e(s, j): e(s, j) \neq \perp \land e(s, j) \in \bigcup_{k \in G \cap R} W_k \land (r_k, c, W_k) \in Q$, then
32.             $W' \leftarrow W' \cup \{(c, e(s, j))\}$
33.         Else
34.             $e \leftarrow (s, j, Empty)$
              $W' \leftarrow W' \cup \{(c, e)\}$
31.     Return $W'$

## Appendix B Leader Election and Group Reconfiguration Protocols

The notations appeared in the leader election protocol and the group reconfiguration protocol follows the same convention listed in Table 3.

The leader election protocol (Algorithm 7) is triggered once the leader is not in the set of live replicas (Line 2). Each triggered replica check whether it satisfies the condition to be the candidate by calling the *SelectLeader* function (Line 8-13). As described in the Leader Election and Group Reconfiguration section, the candidate has the smallest age. If multiple candidates have the same age, the one with the smallest ID is selected.

To achieve state synchrony, the candidate $r_c$ sends the state query message (LE_QUERY) to all live replicas. If a replica has not learned the candidate or has learned a new candidate, the replica will reject the request from $r_c$ by replying the NACK message (Line 17, 21-22). Otherwise, the replica will reply the state query with its state $Q_d$, E, epoch, and configuration (*cid* and *G*). $r_c$, on receiving the states from all live replicas, decides the latest consistent state (Line 47-53) with the following functions.

- *Longest*($\{Q_{d,i}\}$): Selects the longest $Q_d$ from all replicas.
- *Merge*($\{E_i\}$): Returns the union of the Decision sets from all the replicas for all cycles.
- *Latest*($\{(cid_i, G_i)\}$): Returns the largest configuration ID *cid* and the corresponding replica set *G*, which represents the latest configuration seen by the group.
- *Latest*($\{epoch_i\}$): Returns the largest *epoch* which represents the latest leader election seen by the group.

Moreover, in case of any unfinished group reconfiguration, additional states (including the time of the first cycle $t_0$, the start time of the first event from all senders $\{t_{start,s} \mid s \in S\}$, the current state of the application and the corresponding delivered sequence of the event $\lambda_c$, the age of replicas, and the sender set (*S*) are synchronized from $r_c$ to new replicas for state initialization. After receiving the LE_STATE message from $r_c$, the replicas update their state to the decided value. Finally, all replicas load the $r_c$ as the new leader and update the epoch by one.

The uniform agreement on the end of leader election (57-61) ensures that epoch number is monotonically increasing for all sequentially terminated leader elections.

| Algorithm 7. Leader Election |
|---|
| 1.  <u>On any replica:</u> |
| 2.  Upon $r_L \notin R \wedge r_c \neq self$ |
| 3.     *LE ← true* |
| 4.     $r_c$ ← *SelectLeader*() |
| 5.     If $r_c = self$ then |
| 6.        Reliably broadcast LE_QUERY to $G \cap R$ |
| 7.  |
| 8.  *SelectLeader*() |
| 9.     $R_c := \{ r_i \mid \forall r \in G \cap R, r_i.age \leq r.age \}$ |
| 10.    If $|R_c| = 1$, then |
| 11.       Return $r_i: r_i \in R_c$ |
| 12.    Else |
| 13.       Return $r_i: r_i \in R_c \wedge r_i.ID = Min\{r_j.ID \mid r_j \in R_c\}$ |
| 14. |
| 15. <u>On replica $r_i$:</u> |
| 16. Upon LE_QUERY from $r_c$ |
| 17.    If $r_c \in R \wedge r_c = SelectLeader$(), then |
| 18.       *LE ← true* |
| 19.       LE_STATE ← ($Q_d$, E, *cid, G, epoch*) |
| 20.       Reliably send LAST_STATE to $r_c$ |
| 21.    Else |
| 22.       Reliably send NACK to $r_c$ |
| 23. |
| 24. Upon LOAD_LEADER($Q_d'$, E', *epoch', cid', G', Init*) |
|      from $r_c \wedge r_c \in R \wedge r_c = SelectLeader$() |
| 25.    ($Q_d$, E, *cid, G, epoch*) ← ($Q_d'$, E', *cid'*, **G'**, *epoch'* + 1) |
| 26.    ($r_L$, $r_c$) ← ($r_c$, ⊥) |

| | | |
|---|---|---|
| 27. | LE ← false : $r_i \neq r_c$ | |
| 28. | If *newReplica* = *true*, then | // New replica initialization is |
| 29. | Initialize(*Init*) | needed, in case that a group |
| 30. | *newReplica* ← *false* | reconfiguration is interrupted |
| 31. | Reliably send ACK to $r_c$ | by a leader election |
| 32. | | |
| 33. | <u>On leader candidate $r_c$:</u> | |
| 34. | Upon NACK from $r_i$ | |
| 35. | If *self* = *Selectleader*() | |
| 36. | Reliably send LE_QUERY to $r_i$ | |
| 37. | Else | |
| 38. | $r_c ← \bot$ | |
| 39. | | |
| 40. | Upon LE_STATE($Q_{d,i}$, $E_i$, $cid_i$, $G_i$, $epoch_i$) from $r_i$ | |
| 41. | *Events* ← *Events* ∪ {$Q_{d,i}$} | |
| 42. | *Decisions* ← *Decision* ∪ {$E_i$} | |
| 43. | *Configs* ← *Configs* ∪ {($cid_i$, $G_i$)} | |
| 44. | *Epochs* ← *Epochs* ∪ {$epoch_i$} | |
| 45. | *Senders* ← *Senders* ∪ {$r_i$} | |
| 46. | | |
| 47. | Upon $G \cap R \subseteq Senders$ | |
| 48. | $Q_d$ ← *Longest*(*Events*) | |
| 49. | $E$ ← *Merge*(*Decisions*) | |
| 50. | ($cid$, $G$) ← *Latest*(*Configs*) | |
| 51. | *epoch* ← *Latest*(*Epochs*) | // *state*: current application |
| 52. | *Init* ← ($t_0$, {$t_{start,s}$ \| $s \in S$}, ($\lambda_c$, *state*), {($r_i$, $r_i$.*age*) \| $r_i \in$ | state |
| | $G_T$}, $S$) | |
| 53. | LOAD_LEADER ← ($Q_d$, $E$, *epoch*, *cid*, $G$, *Init*) | |
| 54. | Reliably broadcast LOAD_LEADER to $G \cap R$ | |
| 55. | Broadcast $G$ to $S$ | |
| 56. | | |
| 57. | Upon ACK from $r_i$ | // New leader will change the |
| 58. | *Acks* ← (*ACK*, $r_i$) | state only after all live |
| 59. | | replicas have change the |
| 60. | Upon $G \cap R \subseteq$ {$r_i$ \| (*ACK*, $r_i$) $\in Acks$} | state, for uniform agreement |
| 61. | LE ← false | on the end of leader election. |

The group reconfiguration protocol (Algorithm 8) is similar to the leader election protocol, except that it has lower priority, which is reflected by the precondition of checking the flag *LE* in all message handling procedures (Line 11, 19, 30). Group reconfiguration is triggered, when new replicas are added in the survival (i.e., $R \setminus G \neq \emptyset$). $G_T$ caches the latest triggered reconfiguration to preclude any un-necessary re-triggering (Line 2-3). At the end of the reconfiguration, each replica updates the age of all replicas by one (Line 26-27).

| Algorithm 8. Group Reconfiguration | |
|---|---|
| 1.  <u>On any replica:</u> | |
| 2.  Upon $R \setminus G \neq \emptyset \wedge R \neq G_T \wedge LE = false$ | |
| 3.  $G_T ← R$ | |
| 4.  $GR$ ← true | |
| 5.  If $r_L$ = *self*, then | |
| 6.  $cid ← cid + 1$ | |
| 7.  GR_QUERY ← (*epoch*, *cid*) | |
| 8.  Reliably broadcast GR_QUERY to $G_T \cap R$ | |
| 9. | |
| 10.  <u>On replica $r_i$:</u> | |
| 11.  Upon GR_QUERY(*epoch'*, *cid'*) from $r_L \wedge epoch' \geq$ | // Use *cid* to discard messages from a |
| $epoch \wedge cid' > cid \wedge LE = false$ | previous unfinished GR; |
| 12.  $GR$ ← *true* | // *epoch'* ≥ *epoch*: for new members |
| 13.  $cid ← cid'$ | // *cid'* > *cid*: because the new *cid* has not |
| 14.  If *epoch* = 0, then | been received |
| 15.  *epoch* ← *epoch'* | |
| 16.  GE_STATE ← ($Q_d$, $E$, *cid*, *epoch*) | |
| 17.  Reliably send GE_STATE *to* $r_L$ | |
| 18. | |

```
19.  Upon LOAD_CONFIG(Q_d', E', epoch', cid', G_T, Init)
        from r_L ∧ epoch' = epoch ∧ cid' = cid ∧ LE = false
20.      (Q_d, E) ← (Q_d', E')
21.      G ← G_T
22.      LE ← false
23.      If newReplica = true, then
24.          Initialize(Init)
25.          newReplica ← false
26.      For each r ∈ G ∩ R,
27.          r.age ← r.age + 1
28.
29.  On leader r_L:
30.  Upon GE_STATE(Q_d,i, E_i, cid_i, epoch_i) from r_i ∧ epoch_i
        = epoch ∧ cid_i = cid ∧ LE = false
31.      Events ← Events ∪ {Q_d,i}
32.      Decisions ← Decision ∪ {E_i}
33.      Senders ← Senders ∪ {r_i}
34.
35.  Upon G_T ∩ R ⊆ Senders
36.      Q_d ← Longest(Events)
37.      E ← Merge(Decisions)
38.      Init ← (t_0, {t_start,s | s ∈ S}, (λ_c, state), {(r_i, r_i.age) | r_i ∈      // state: current application state
             G_T}, S)
39.      LOAD_CONFIG ← (Q_d, E, epoch, cid, G_T, Init)
40.      Reliably broadcast LOAD_CONFIG to G_T ∩ R
41.      Broadcast G_T to S
42.      G_T ← ∅
```

# Appendix C Proposition Proofs

**Lemma 6.1 (Leader Election Synchrony).** All the live replicas in $G$ maintain the same $Q_d$, $E$, and G after a leader election.

**Proof.** First, only one leader will eventually be elected by all the live replicas. It can be inferred by two cases. In the first case, the group is not partitioned. Then all the live replicas know each other, and the *SelectLeader* function ensures that only one leader is elected by all live replica. In the second case, the group is partitioned. Without loss of generality, suppose there are two different leaders, denoted by $r_{L,1}$ and $r_{L,2}$. $r_{L,1}$ is elected by replica set $P$ and $r_{L,2}$ is elected by replica set $Q$. $r_{L,1} \notin Q$, $r_{L,2} \notin P$, and $P = G \setminus Q$. Following the partial synchrony assumption, if the replicas in P never know $Q$ and the vice versa, then either $P$ or $Q$ is removed by the Rendezvous of the group. Following the assumption that there is only one Rendezvous for each replica group, then only one partition, either $P$ or $Q$, will eventually survive. Thus, eventually there is only one leader, either $r_{L,1}$ or $r_{L,2}$ is the leader of the group.

When a new leader is elected by all replicas, it will determine the $Q_d$, $E$, and $G$ and broadcast them to all the live replicas. Through the reliable underlying channel, all replicas will eventually load the same $Q_d$, $E$, and $G$ after leader election. Moreover, a monotonic epoch number is used to avoid a replica load the state from an old leader. Thus, all live replicas will eventually load the same $Q_d$, $E$ and $G$ after the leader election of the largest epoch. □

**Lemma 6.2 (Group Reconfiguration Synchrony).** All the live replicas in $G$ maintain the same $Q_d$, $E$, and $G$ after a group reconfiguration.

The proof of group reconfiguration synchrony is the same as that of leader election synchrony. Thus, it is not repeated here.

**Lemma 6.3 (Consensus Agreement).** If a live replica $r_i \in G$ delivers an event $e$ to $E(c)$ from a consensus instance for cycle $c$, then $e$ is eventually delivered to $E(c)$ by all the live replicas.

**Proof.** If there is a leader election or a group reconfiguration before the consensus instance terminates, then Lemma 6.1 and Lemma 6.2 ensure that all replicas will have $e$ in $E(c)$. If there is no leader election or group reconfiguration before the consensus instance terminates. The reliable underlying communication channel ensures that all the live replicas will eventually receive the same decision from the leader. Since $r_i$ had delivered $e$ into $E(c)$, $e$ is the in the decision for cycle $c$. Therefore, $e$ will be eventually received and delivered by all live replicas.

<div align="right">□</div>

With the above lemmas, the main result can be obtained. But before it, an important property of the late event handling approach needs to be verified first.

**Lemma 6.4 ($\Omega(s, c)$ Agreement).** All the live replicas in $G$ expect to deliver the same set of events $\Omega(s, c)$ for sender $s \in S$ and cycle $c$.

**Proof.** The lemma can be proved by induction.

Basis Step: when $c = c_0$, i.e., the cycle of receiving the first event from $s$ based on $t_{recv,s}(1)$ , then $\Omega(s, c) = \{e(s, 0)\}$.

Induction Step: Assume all the live replicas in $G$ expect to deliver the same set of events $\Omega(s, c_k)$ for sender $s \in S$ and cycle $c_k$ ($c_k \geq c_0$). Then, for cycle $c_k + 1$, there are two cases for discussion.

1. If there is no consensus instance for cycle $c_k$, then $MaxSeq(s, c_k) = Seq(s, c_k)$ and $\Omega(s, c_k + 1) = \{e(s, c_k + 1)\}$ on all replicas.
2. If there is consensus instance for cycle $c_k$, then following Lemma 6.3, all live replicas will eventually deliver the same events to $E(c_k)$. Let $Seq(s, j)$ be the maximal sequence number of non-empty events in $E(c_k)$. Then, $MaxSeq(s, c_k) = Seq(s, j)$, and $\Omega(s, c_k + 1) = \{e(s, j + 1), e(s, j+2), \ldots, e(s, c_{k+1})\}$ on all the live replicas.

By the principle of mathematical induction, it follows that the lemma is true for all cycles after $c_0$.

<div align="right">□</div>

**Theorem 6.5 (Total-order Event Delivery).** If a live replica $r_i \in G$ delivers two different events $e_1$ and $e_2$ into $Q_d$ with $\lambda_1$ and $\lambda_2$, then $e_1$ and $e_2$ will eventually be delivered into $Q_d$ on all the live replicas with $\lambda_1$ and $\lambda_2$ being two non-negative integer numbers and $\lambda_1 \neq \lambda_2$.

**Proof.** Since all replicas share the same sender set $S$, Lemma 6.3 and Lemma 6.4 ensures that all the live replicas will eventually deliver the same set of events for any cycle, either directly from received events (Line 5 - 14 of Algorithm 1) or from the consensus result (Line 2 - 3 of Algorithm 1).

Let $e_1(s_1, j_1)$ and $e_2(s_2, j_2)$ be delivered on r for the cycle $c_1$ and cycle $c_2$. If $c_1 = c_2 = c$, then $(c, \gamma_1, e_1)$ and $(c, \gamma_2, e_2)$ will be eventually delivered into the $Q_d$ of all replicas. If $c_1 \neq c_2$, then $(c_1,$ $\gamma_1$, $e_1$) and ($c_2$, $\gamma_2$, $e_2$) will be eventually delivered into the $Q_d$ of all replicas. Moreover, since $\gamma_1$ and $\gamma_2$ are determined only by $s_1$, $s_2$, $j_1$, and $j_2$, $\gamma_1 \neq \gamma_2$ for different $e_1$ and $e_2$. Since $Q_d$ is linearly ordered by $c$ and then by $\gamma$, there exists a mapping from each unique ($c$, $\gamma$) to a unique non-negative integer number $\lambda$ and let $\varphi(c, \gamma) = \lambda$ be such mapping function. Let $\varphi(c_1, \gamma_1) = \lambda_1$ and $\varphi(c_2, \gamma_2) = \lambda_2$. Then, all replicas will eventually deliver ($\lambda_1$, $e_1$) and ($\lambda_2$, $e_2$) and $\lambda_1 \neq \lambda_2$.  □

**Corollary 6.6 (Replica Synchronization).** All the live replicas in $G$ maintains the same state of their virtual objects.

Corollary 6.6 can be directly inferred from Theorem 6.5.

**Theorem 6.7 (Garbage Collection Safety).** If event $e$ is removed from $Q_d$ on $r_i \in G$, then $e$ has been handled by the application on all the live replicas in $G$.

**Proof.** Theorem 6.5 ensures that if $e$ is in the $Q_d$ of $r_i$, then e is or was in the $Q_d$ of all the live replicas in $G$ with the same $\lambda$. In Algorithm 3, if $e$ can be removed from $r_i$, then $r_i$ must have received $\lambda_c$'s at least equal to $\lambda$ from all the live replicas. Since events are delivered to the application in sequence, $e$ must have been delivered to the application on all replicas.  □

**Corollary 6.8 (Total-order Event Delivery with Sender Join).** All the live replicas in $G$ deliver the same first event $e_0$ from a neighbor $s$ with the same delivery sequence $\lambda_0$.

**Proof.** Theorem 6.5 ensures that the ADD_NEIGHBOR event is delivered to the $Q_d$ of all replicas with the same $\lambda$. Since ($\lambda$, $e$) and ($c$, $\gamma$, $e$) has a one-to-one mapping for the same event, all replicas deliver ADD_NEIGHBOR for the same cycle. Moreover, since, $n$, $\Delta t$, are fixed, all replicas are timed to deliver the first event $e_0$ from $s$ for the same future cycle $c_k$. Theorem 6.5 ensures that $e_0$ is delivered with the same delivery sequence $\lambda_0$ on all live replicas.  □

**Corollary 6.9 (Total-order Event Delivery with Sender Leave).** All the live replicas in $G$ deliver the same last event $e_\infty$ from a neighbor $s$ with the same delivery sequence $\lambda_\infty$.

**Proof.** Theorem 6.5 ensures that the RM_NEIGHBOR event is delivered to the $Q_d$ of all replicas with the same $\lambda$. Since ($\lambda$, $e$) and ($c$, $\gamma$, $e$) has a one-to-one mapping for the same event, all replicas handle RM_NEIGHBOR for the same cycle $c$. From cycle $c + 1$, $s$ will be removed from the $S$. Thus, all replicas will deliver the last event $e_\infty$ of $s$ at $c$. Theorem 6.5 ensures that e$_\infty$ is delivered with the same delivery sequence $\lambda_\infty$ on all live replicas.  □